\documentclass[final,5p,twocolumn]{elsarticle}

\usepackage{amssymb}
\usepackage{hyperref}
\hypersetup{colorlinks=true,citecolor=blue}
\usepackage{graphicx,amsmath}
\usepackage{ifthen}
\usepackage{endnotes}
\usepackage{multirow}
\def\beq{\begin{equation}}
\def\eeq{\end{equation}}
\def\bea{\begin{eqnarray}}
\def\eea{\end{eqnarray}}

\biboptions{round,comma,authoryear}

\journal{Energy Policy}

\begin{document}

\begin{frontmatter}


\title{On the global economic potentials and marginal costs of non-renewable resources and the price of energy commodities}
\author{Jean-Fran\c{c}ois Mercure \corref{cor1} \fnref{fn1}}
\ead{jm801@cam.ac.uk}
\cortext[cor1]{Corresponding author: Jean-Fran\c{c}ois Mercure}
\fntext[fn1]{Tel: +44 (0) 1223337126, Fax: +44 (0) 1223337130}
\author{Pablo Salas}
\address{Cambridge Centre for Climate Change Mitigation Research (4CMR), Department of Land Economy, University of Cambridge, 19 Silver Street, Cambridge, CB3 1EP, United Kingdom}

\begin{abstract}

A model is presented in this work for simulating endogenously the evolution of the marginal costs of production of energy carriers from non-renewable resources, their consumption, depletion pathways and timescales. Such marginal costs can be used to simulate the long term average price formation of energy commodities. Drawing on previous work where a global database of energy resource economic potentials was constructed, this work uses cost distributions of non-renewable resources in order to evaluate global flows of energy commodities. A mathematical framework is given to calculate endogenous flows of energy resources given an exogenous commodity price path. This framework can be used in reverse in order to calculate an exogenous marginal cost of production of energy carriers given an exogenous carrier demand. Using rigid price inelastic assumptions independent of the economy, these two approaches generate limiting scenarios that depict extreme use of natural resources. This is useful to characterise the current state and possible uses of remaining non-renewable resources such as fossil fuels and natural uranium. The theory is however designed for use within economic or technology models that allow technology substitutions. In this work, it is implemented in the global power sector model FTT:Power. Policy implications are given.

\end{abstract}


\end{frontmatter}


\section{Introduction}

\subsection{Energy-Economic-Environmental interactions}

The use of large scale models for exploring Energy-Economic-Environmental (E3) interactions is crucial for devising policy aimed at addressing coupled economic and environmental problems and achieve related policy goals. This is due to the fact that in such complex and highly correlated systems, while conceptual difficulties arise in attempting to understand the systems-wide influence of individual policies and regulations, significant complications arise in the potential mutual influence between several such policies \citep{IPCCAR4Ch11}. This includes for instance the strong interaction between government support for novel transportation technology and power sector or land use management, and their very uncertain effect on global emissions, which depend highly on their timing, technology diffusion timescales and energy conversion efficiencies \citep[as examples of differences in estimations of potential emissions reductions for the transport sector, see][]{vanVliet2010,vanVliet2011,Pasaoglu2012,Takeshita2011,Takeshita2011b}. It has been widely recognised that large expansions in modelling capacity are required in order to better predict the likely result of comprehensive policy portfolios, which should include combinations between top-down economic models and bottom-up technology models \citep[see for instance][]{Kohler2006b,Kohler2006,Grubb2002}. While common economic models can evaluate the global demand for energy, transport, materials, goods and services, they generally do not represent with much detail the way in which their supply is produced and at which costs, from lack of technology resolution, or none altogether. This generates for instance significant uncertainty over production efficiency, carbon intensity and greenhouse gas (GHG) emissions. Meanwhile, bottom-up technology models generally take demand values (energy, services, goods, etc) as given and therefore, although they are able to generate prices and accurate efficiency values and emissions factors, they do not capture the interaction between prices and demand \citep[for details on these aspects for several existing models, see][]{Edenhofer2006,Edenhofer2010}). Coupling bottom-up and top-down models generates the most powerful method to capture systems-wide and economy-wide coupled interactions, which are currently strongly required for devising sensible climate change mitigation policy \citep{Kohler2006b}.

Energy flows, originating from natural resources, are a necessary component for all sectors of the world economy. Although the economic output of the energy sector accounts only for a small fraction of the world gross domestic product (GDP),\endnote{The global output of the energy and fuel supply industries makes 2-3\% of global GDP and decreasing, values obtained from our own E3MG-FTT calculations \cite{Mercure2013a}.} changes in the prices of energy carriers have pronounced consequences on the output of most other economic sectors \citep[see for instance][]{Jones2004}).\endnote{This is also  a pronounced effect in E3MG-FTT results.} Since the price of energy carriers are reflected in the prices of goods and services originating from energy intensive sectors, such changes can lead to increased inflation, decreases in economic output and reduced paces of economic development.  Many attempts have been made to capture such interactions between energy, the economy and the environment in computer models \citep[see for instance][ and the various models reviewed]{Edenhofer2006,Edenhofer2010}.  While many models of E3 interactions do not incorporate explicit representations of natural resource use and depletion, or the physical limits to available energy flows, very few feature endogenous exploitation costs of non-renewable resources and none of them to our knowledge feature a particular emphasis on uncertainty in the economic or technical potentials of natural resources.\endnote{Most models rely on outdated and fixed (i.e. not time dependent) cost-supply curves from \cite{Rogner1997}.} For this reason, in previous work we defined a theoretical framework and built an extensive database with a resolution of 190 countries for limiting and tracking the use of natural resources in models of global energy systems \citep{MercureSalas2012}, which, although adaptable to any energy systems modelling framework, was designed for use in the model Future Technology Transformations in the Power sector (FTT:Power) \citep{Mercure2012}. FTT:Power is based on a theoretical framework for technology diffusion \citep{Mercure2012b,Mercure2013b}, integrated as a bottom-up component to the Energy-Economy-Environment model at the Global level \citep[E3MG, for descriptions see][]{E3MG, Barker2012, Barker2006, Barker2010,Kohler2006b}.

Modelling energy systems realistically requires the representation of many complex interactions between different types of systems, which must respond to the economic climate and natural environment at every point in time. This involves modelling the behaviour of actors who influence the working and composition of the technological mix within the system. Once this mix is defined, the requirements in terms of energy resources are straightforward to evaluate. Global energy demand is strongly influenced by the price of energy carriers,\endnote{As can readily be observed using E3MG-FTT with different scenarios of energy prices. E3MG-FTT is an econometric model that extrapolates such trends from data \citep{Mercure2013a}.} generating a feedback interaction between the economy and the global energy system through demand and prices \citep{Mercure2013a}. Meanwhile, the cost of energy resources influences the choice of investors in energy systems and thus the technology composition, as well as the cost of producing energy carriers. Therefore, a second strong feedback interaction exists between the global energy system and the natural environment through the exploitation of resources through demand and prices. As described earlier by one of us \citep{Mercure2013b,Mercure2012, Mercure2012b}, the evolution of technology in most sectors, including the power sector, is well described by a coupled family of non-linear differential equations that simulates transitions between energy technology systems, changes that are driven by the trend of investor decisions, an approach supported by an extensive empirical literature \citep[see for instance][]{Grubler2012, Marchetti1978, Grubler1999, Wilson2009, Bass1969, Sharif1976,Bhargava1989,Morris2003, Grubler1990}. Meanwhile, the cost of producing energy carriers is influenced by that of natural resources, as well as and through components such as investment, maintenance, capacity factors and taxes or carbon costs, all of which should be considered when calculating the cost of electricity production, for which for instance the Levelised Cost of Electricity \citep[the LCOE, see for instance][]{IEAProjCosts}, in the case of the power sector, is a good representation of the way investors evaluate technology costs (and in a similar construction for other sectors of technology). As argued in our previous work \citep{Mercure2012,MercureSalas2012}, the limitation of the expansion of certain types of energy systems is well reproduced by cost-supply curves, which track the increasing of the marginal cost of production of energy with increasing demand, through its influence into certain components of the LCOE (e.g. fuel costs, capacity factors, investment costs, etc). 

Modelling energy flows from renewables and non-renewable resources entails large conceptual differences. Cost-supply curves have been generated for different types of renewable resources in works by \cite{HoogwijkThesis, Hoogwijk2004, Hoogwijk2005, deVries2007, vanVuuren2009}, using the cumulative sum of cost rankings of the global number of potential sites for energy production by type (wind, solar and biomass energy). This involves the assumption that these renewable resources are chosen and exploited in order of cost, starting with the most profitable development ventures. The cost-quantity availability of non-renewable resources such as oil and gas can also be expressed using a cost-quantity curve \cite[as in][]{Rogner1997, MercureSalas2012}, which expresses a quantity available at a certain exploitation cost rather than a flow. Such a curve, however, is much more difficult to interpret in order to derive marginal costs, since taking the assumption that consumption progresses in perfect order of exploitation cost is not reliable, and the gradual depletion of fixed amounts of resource means that the cost-quantity curve changes with time. In contrast, as apparent in the oil industry for instance, the exploitation costs of existing projects cover a wide range rather than a single competitive value, depending on the nature and quality of resource occurrences \citep{IEAETSAP2010,IEAETSAP2010b}. This range is determined by the price of oil. There is thus a connection between the supply and the price of energy commodities, where higher prices enable production at higher costs, and therefore the accession of larger amounts of resource at such costs. Meanwhile, the demand for energy commodities may justify  increases of prices, in order to enable production at higher costs, such that the demand is met by the supply, using ever more difficult and expensive methods, locations and types of resources (ultra-deep offshore drilling, arctic sites, tar sands, oil shales, etc). However, high prices, as for instance generated by depletion and scarcity, may also be avoided by simply phasing out the use of certain types of high price commodities, replacing them by other types. Such substitutions actually stem from technology substitutions, which can become economical in the event of the price of some commodities increasing (e.g. replacing oil by coal for producing electricity, which entails phasing out existing technology, and therefore cannot be performed instantaneously\endnote{Power sector technology substitution can occur at timescales not much shorter than 30-40 years.}). However, technology lock-ins can also impede sector transformations that would enable avoiding price escalations. Thus, technology substitution dynamics have a strong long term influence as well on energy commodity prices. And conversely, the price of energy commodities has a strong long term influence on the technology mix.

\subsection{Oil price models}

Many models aiming at describing the dynamics of the price of oil exist in the literature \citep[for instance][]{Rehrl2006,Reynolds1999,Carlson2007,Gallo2010,Michl2007}. Models are mostly classified into two groups, based on whether they use the Hotelling Principle or the Hubbert Peak approach \citep{Reynolds2012}. The Hotelling Principle generates the optimal extraction rate of a known non-renewable resource, where its price follows a rate of increase equal to either a social discount rate or an interest rate \citep{Hotelling1931,Perman2003}. Meanwhile, the Hubbert Peak theory is an empirical observation that Hubbert made on historical US oil production data from 1900 to 1962 using a logistic trend \citep{Hubbert1956,Hubbert1962}, for which a theoretical derivation was later developed, based on probabilistic arguments concerning the rate of oil field discoveries, that generates the logistic (or more general) mathematical form of the observed trend. In this theory, while random drilling generates a rate of discovery which is proportional to the amount of undiscovered oil left in a geographical area, oil can statistically more easily be found near existing fields, generating a quadratic term in the probability function (the `information effect'), leading to a logistic differential equation \citep{Reynolds1999,Rehrl2006}. Both theoretical approaches can be criticised, for different reasons described below.

While the Hotelling Principle does provide the optimal rate of resource extraction given a certain resource base, it is very unlikely that real resource extraction activities follow a path anywhere near optimal, for the following reasons. As \cite{Norgaard1990} argues (using the `Mayflower problem'), mineral extraction firms would need to know the total amount, location and quality of resources in the ground over which they have contracts, and would need to have perfect foresight, both required to enable them to figure out how to follow the optimal pathway prescribed by Hotelling \cite[see also][]{Reynolds2012}. There is no clear empirical evidence in the literature to justify the assumption that energy or mining firms actually follow such patterns \citep[see for instance the textbook][pp. 527-532]{Perman2003}. 

Meanwhile, the extended Hubbert Peak theory generates a supply that is entirely independent of demand, based solely on the rate of discoveries. Independent and individual Hubbert peaks have been observed for different types of oil occurrences \citep[conventional US oil fields, deepwater, Alaska, as described in][]{Rehrl2006}. However, it has also been shown that OPEC behaviour does not follow a logistic trend and has a reserve to production ratio of about 80~years, higher than the rest of the world \citep[see below in section \ref{sect:restoprod}, as well as in][]{Rehrl2006}. Thus it does not apply to all situations, and if it did, it would mean that oil consumption would follow a sum of rigid individual logistic functions, independent of oil demand, or alternatively with the demand and supply independent of the oil price.

In contrast to this, for example, as the price of oil passed the economic threshold of 85-95\$/boe \citep{NEB2011} in early 2007, intense activity started in the Canadian tar sands, which subsequently came to a standstill slightly later in 2008 when the price dipped below that value again, and started again later in 2009 when the price increased again above a similar threshold \citep[see for instance in][]{Earthworks2010,IEAOilInfo2011}.\endnote{This can be observed from data with a dip in oil sands production at the dip in crude oil price in IEA oil price and production data around 2008-2009 \citep{IEAOilInfo2011}.} This indicates how some exploitation activities occur very near the competitive margin, the production of which can be turned on or off, following changes in demand rather than a Hubbert peak. Thus, the difference between a price dependent supply or demand and a sum of rigid Hubbert cycles is buffered by high cost resources at the edge of profitability and/or by OPEC monopolistic behaviour. \cite{Sorrell2010} states that `Most of the world's conventional oil was discovered between 1946 and 1980 and since that time annual production has exceeded annual discoveries', indicating that a significant fraction of the original oil resources in place have already been discovered. These are not necessarily under exploitation, and therefore discoveries and changes in the knowledge about the location of oil fields is apparently not the single determinant of oil demand and supply. Finally, Hubbert peak theory ignores the dynamics of technological change and technology substitution, which influences the demand by transiting away from expensive resources. We thus argue in this paper that neither of these two strict approaches are appropriate to project energy carrier prices, and propose an alternative model that does not assume knowledge about resources, perfect foresight, or that is based strictly on rates of discoveries.

\subsection{New approach to modelling carrier flows and prices}

In this paper, we present a detailed description of a theoretical framework to treat the exploitation of non-renewable resources (stock resources henceforth) and the prices of their associated energy commodities. This model is designed to be used in conjunction with a database for natural resources as well as with a model of technology substitution. It is a general model that can be used to represent any type of  natural resource consuming systems, but is applied here within our power sector model FTT:Power-E3MG. This model does not, however, treat the effect on prices of speculation and hoarding, or of supply problems related to geopolitical events, making it unable to forecast short term price fluctuations that commonly occur in global markets.\endnote{It does not replicate the reasons why people hoard stockpiles or the expectations of political conflicts or supply restrictions related to political decisions. With assumptions over a medium term artificial demand component (positive or negative) originating from changes in stockpiles, the model could be used to predict medium term price fluctuations, but not price hikes associated to very short term supply problems.} It does, however, project inevitable long term base price values below which the production sector cannot supply the demand, and therefore provides a lower bound for price values, equal to the marginal cost of production. It may additionally be argued that hoarding and speculation over futures can only ultimately lead to bubbles and cyclic fluctuations, since (1) speculation and hoarding cannot occur without storage space, and (2) storage space is finite and stored commodities must eventually return to the market for this activity to be profitable (i.e. storage space cannot expand indefinitely). Thus the artificial demand (i.e. demand unrelated to immediate consumption) generated by speculation is cyclic and evens out to zero in the long run, generating price fluctuations that oscillate. Price fluctuations observed occur at faster frequencies than the fastest possible rates of technological change, and are therefore seen as volatility (i.e. quantifiable risk). Therefore it is mostly long term price changes that are truly relevant in energy systems modelling, along with the cost associated to volatility, and thus the omission of hoarding and speculation in this theory is not expected to affect the results significantly, and the cost of volatility can be included into the sum of fixed costs, added to the marginal cost of production. Similarly, only long term prices should be considered in energy and climate policy.

The results of this study provide insight into energy policy-making, particularly in the context of climate change policy. Through the use of limiting assumptions, limits to the use of stock resources, as well limits to the behaviour of commodity prices are explored, providing information on the range of possible futures for the use of stock resources, their prices and the potential scale of global emissions. These scenarios are price inelastic, however, and thus more realistic scenarios are given, obtained when connecting this theory with a model of technological change. While this insight can be used directly in better understanding the current state of global stock energy resources, this model reveals its real value through its integration into FTT:Power-E3MG, and in future work into the full FTT family of technology models interacting with E3MG. 

This paper progresses as follows. We first present a formal definition of a dynamic model for tracking the use and depletion of stock resources based on cost distributions. This model can be used with two types of assumptions, either an endogenous commodity supply given a commodity price (the forwards problem), or an endogenous commodity price given an exogenous demand (the reverse problem). As modelling exercises, the model is used with our natural resource database \citep{MercureSalas2012} in these two limiting assumptions cases to determine, with confidence ranges related to uncertainty in resource assessments, limits to potential future supplies or prices in situations of fixed technology. The results of these two exercises are instructive as they provide a clear picture of the limits to stock resources. However, they do not generate realistic scenarios of resource use, since they use assumptions where prices are independent of the economy and of one another, and demand values are independent of prices. In order to create fully dynamic scenarios that include possible technology substitutions, this model is integrated into FTT:Power. Two commodity supply and price scenarios are given where technology substitutions in the power sector enable the energy system to avoid price escalations. The impacts of these considerations onto climate policy are discussed.

\section{Theoretical framework}

\subsection{Flows from resources to reserves and the consumption of reserves}

\begin{figure*}[t]
	\begin{center}
		\includegraphics[angle = -90, width=1.5\columnwidth]{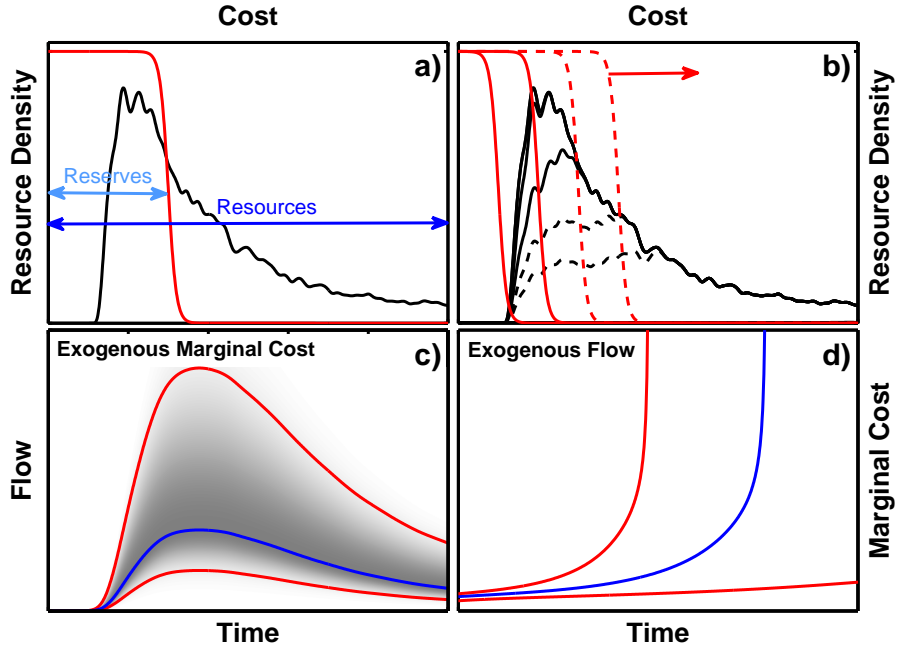}
	\end{center}
	\caption{Process of the gradual non-renewable resource consumption and related price changes. $a)$ Original (current) resource distribution as a function of extraction cost (black curve) and probability function for resource extraction given the price of the resource commodity that delimit reserves out of resources (red curve), its maximum being equal to 1. $b)$ Dynamic process of resource depletion as the price of the resource commodity increases. The black curves corresponds to the distribution of resources left after increasing amounts of time have passed and increasing amounts of resource have been consumed, associated with increasing prices for the energy commodity. $c)$ Flow curve and uncertainty range for the amount of resource unlocked by an increasing exogenous price of the commodity which allows the resource exploitation to proceed up to an upper limit marginal cost (function of time). The area between the red curves indicates the 96\% confidence level region, while the blue curve represents the most probable curve. $d)$ Marginal cost of production of the commodity and its uncertainty range for an exogenous demand (function of time). The values diverge at depletion, which, due to uncertainty, can occur at different values of cumulative production.}
	\label{fig:FlowCurves4All}
\end{figure*}

A fixed energy carrier demand, when met by renewable energy resources, results in fixed levels of use of these resources. However, when met by stock resources, it results in particular rates for their depletion. The gradual depletion of stock resources generates gradual increases in their cost of exploitation, an effect which is due to the natural average tendency towards the exploitation of resource occurrences with lower extraction costs first, and costs increase following their depletion. In our previous work \citep{MercureSalas2012}, we calculated cost-quantity curves for fossil and nuclear resources. When allocating energy demand between all potential energy sources, dynamic rates of exploitation of stock resources emerge, which result in particular lengths of time for these resources to reach depletion, and which depend on the price of the associated energy commodities. We introduce in this section a simple mathematical model that generates a relationship between rates of exploitation of a given finite resource base and the price of the associated commodity. This relationship is not functional however, but as we show, involves hysteresis and irreversibility, and therefore strong path dependence. The supplementary information provides additional mathematical details in order to understand all of its properties, which are not shown here in order not to lose the reader into strongly theoretical considerations.

In a hypothetical world with a perfectly efficient and competitive global energy market, fast rates of resource extraction and no speculation over the future value of these commodities, or monopolistic or cartel price markups, only the resources with lowest extraction costs would be traded at any time. In such a situation, resource consumption would proceed in perfect order of increasing extraction cost (or level of technical difficulty of recovery) and follow closely a cost-quantity curve. This is of course not the case, and in reality, owners of low cost reserves delay their extraction, forcing the exploitation of more expensive resources while low cost resources are not yet depleted.\endnote{This phenomenon is know to take place in the oil market, as noted for instance in \cite{Johansson2009}, where furthermore the behaviour of OPEC is modelled, and this phenomenon is projected to remain present even in scenarios of climate policy.} Thus, the demand for energy commodities is met by resource types with extraction costs within a certain range \citep[see for instance the variety of resource types studied in ][]{IEAETSAP2010,IEAETSAP2010b}, delimited by the price of energy carriers, which determines what is economic and what is not, giving a corresponding range of margins of profit for different resource types. For example, the price of oil determines which of known oil fields are deemed economic to exploit, and the remaining fields are reserved for a future in which a higher price of oil is expected. Additionally, however, the extraction of existing economic reserves is done over a length of time, and some reserves are kept untapped for a future where higher prices are expected. The upper limits for the extraction cost values that are still considered economical given the prices of energy carriers, in other words the cost of the most expensive resource exploited (i.e. the marginal cost), are defined by the differences between the prices of the commodities and the sum of all fixed costs such as those associated to transformation processes and transport, and the minimum profit margins that industries will consider. Increases in the prices of energy carriers enable wider ranges of natural resource types to be exploited, for instance low grade or unconventional fossil fuels, which are not profitable under low price conditions. 

This behaviour can be summarised by stating that increases in the price of energy carriers \emph{unlock} additional energy resources by enabling a higher marginal cost of production to become economical. While low cost stock resources become increasingly depleted, increases in price of energy carriers enable the exploitation of additional high cost resources in order to supply the demand. It may thus be inferred that price paths in time produce supply paths in time for energy resources. It is however the demand for energy carriers that determine their prices: they adjust in such a way that the supply resulting from the sum of resources unlocked meets the price adjusted demand.\endnote{Strictly speaking, this is true over the medium term where demand adjustments have time to take place.}

Following the terminology of \cite{McKelvey1972} and \cite{Rogner1997}, reserves are seen as continuously consumed but also continuously expanded at the expense of resources. Reserves are defined as those currently economical to exploit, and the boundary delimiting reserves from resources is defined by prices, which evolves in order to maintain a certain level of reserves out of existing resources, with respect to the demand. As we show below, it is remarkable that on the global level, the ratio between the rate of consumption of oil and gas to the size of their associated reserves has been nearly constant in recent history. This aspect strongly strengthens the assertion that it is the evolution of prices that enables the size of reserves to follow the magnitude of their respective consumption levels, which we take as a starting postulate in order to define the theory given here.

Meanwhile, reserves can expand with discoveries and technological change. Discoveries of resources that are not in areas where they are expected to be found (i.e. not counted as \emph{inferred} or \emph{speculative} resources) are part of the unknown/unknowable resource endowment, and therefore may be considered as uncertainty. Similarly, technological breakthroughs in the oil industry are not straightforward to predict, and are also considered part of the uncertainty. A complete description of our approach to uncertainty is given in \cite{MercureSalas2012}. A Monte-Carlo simulations approach may be used by defining large numbers of variations of cost distributions of resources actually in-place, whether discovered or not, and generating different scenarios of resource use for different levels of scarcity. Therefore, in the model presented here, the process of discovery is not assumed to take an important role, or, in other words, the process of discovery is considered part of the gradual resource consumption process.

\subsection{Calculation of energy flows from existing stocks \label{sect:FlowEqn}}

Flows of stock resources and associated depletion can be calculated given time paths for their associated carrier prices which \emph{unlock} just the right amount of energy at every time step that is not already produced by other resources types, in order to supply the total energy demand. This increase in price is associated with the marginal cost of production for this energy resource. 

Following the first panel of figure~\ref{fig:FlowCurves4All} \citep[see also figure 1 in ][]{MercureSalas2012}, we take $n_0(C)$ as the initial cost distribution of a particular type of resource (e.g. oil, gas, coal or uranium), and a time dependent resource distribution $n(C,t)$ which represents the cost distributed amounts of resource left at time $t$. $n(C,t)$ is equivalent to a histogram of all  units (e.g. barrels, tons, etc) of a particular resource type that are assumed to be in place and ranked according to their cost of exploitation. We take the assumption that the rate of extraction of stock resources in each cost range (i.e. between $C$ and $C+dC$), at any time and price, is proportional to the amount of resource left in that cost range, if the latter is considered economical to extract, with a probability that it is considered so.\endnote{For example, the supply is proportional to the number and/or size of wells or mines, which is proportional to the amount of economical resources (reserves) left.} We take a continuous step-like function $f(P(t)-C)$,\endnote{The function can be a smooth rounded step to reflect variations over the response to the price, see the supplementary information for details. This is similar to particular conceptual problem in reaction chemistry and physics, see \cite{Mercure2005}.} which equals one below this boundary and zero above, as the probability of extraction of resources in the cost between $C$ and $C+dC$. This is shown in panel $a)$ of figure~\ref{fig:FlowCurves4All}, where a hypothetical resource distribution is shown as a solid black line, $f(P(t)-C)$ is shown as a red curve that converges to one towards low values of $C$, and reserves correspond to the section of the distribution situated to the left of this curve.

While the remaining cost distributed resources are $n(C,t)$,\endnote{The $total$ remaining resources correspond to the integral of $n$, $\int_0^\infty n(C,t) dC$. } cost distributed reserves correspond to $n(C,t)f(P(t)-C)$. If the constant fraction $\nu_0$ of reserves in each cost range are extracted at any time, the flow of resources is therefore as follows:
\beq
{dn(C,t)\over dt} = -\nu_0 n(C,t) f(P(t)-C),
\label{eq:ResDep}
\eeq
The time dependent energy carrier price $P(t)$ drives the evolution in cost of the boundary between reserves and resources. For a constant or increasing commodity demand, as the size of reserves decreases following consumption, this flow decreases and generates an upwards movement of $P(t)$, shown in panel $b)$, where the distribution of resources decreases in the low cost range and the boundary $f(P(t)-C)$ moves to the right. This produces a time dependent supply (or flow) $F(t)$ which is the sum of the production in all economical extraction cost ranges during the unit time $dt$:
\beq
F(t) = - {dN(t)\over dt} = - \int_0^{\infty} {dn(C,t) \over dt} dC,
\label{eq:IntResDep}
\eeq
providing a connection between commodity prices $P(t)$ and commodity flows $F(t)$. Note however that this connection does not have a unique functional form, but changes depending on the history of $P(t)$ and the amount of resources \emph{left}, a fact due to the integral of eq.~\ref{eq:IntResDep}, where for instance completely different historical paths and values of $P$ at a particular time can lead to the same value of $F$. This indicates possible hysteresis and corresponds to path dependence. Details of these properties are given in the supplementary information.

These equations can be used in two ways, depending which variable is taken as exogenous and which is endogenous. For an assumed time dependent price $P(t)$, the flow $F(t)$ is straightforward to calculate numerically using a discrete time step using equations~\ref{eq:ResDep} and~\ref{eq:IntResDep} (the forward problem). This is shown in panel $c)$, where a range of time dependent flows is depicted, associated with the uncertainty in the amount of resources that actually turn out to be in place, using a linearly increasing commodity price. This flow is low at low price values, where no resources are economical to exploit, and low at high prices, where all resources have been consumed. 

Conversely, for an assumed commodity demand, the price value that \emph{unlocks} just the right amount of resources can be obtained by trial and error using an optimisation procedure (the reverse problem).\endnote{This is done by trial and error because it is not possible to know the integrand of eq.~\ref{eq:IntResDep}, which is function of $C$, from its integrated result $F(t)$. One must therefore take guesses over which value of $P(t)$, $f(P(t)-C)$ and therefore of the integrand, that gives a particular value for $F(t)$.} Panel $d)$ depicts schematically the result of such a calculation with a range associated with the uncertain amount of resources in place, where lower prices result from higher amounts resources and vice versa, using as an example a large constant commodity demand. This results in a price value that gradually increases as resources are consumed, accelerating when remaining resources are small compared to the level of consumption, forcing the exploitation of expensive resource occurrences. It eventually diverges as depletion sets in. In both examples, the red curves delimit the 96\% confidence region associated with resource assessment uncertainty, while the blue curve denotes the most probable values.

\subsection{Constant global production to reserve ratios\label{sect:restoprod}}

\begin{figure*}[t]
	\begin{minipage}[t]{1\columnwidth}
		\begin{center}
			\includegraphics[width=1\columnwidth]{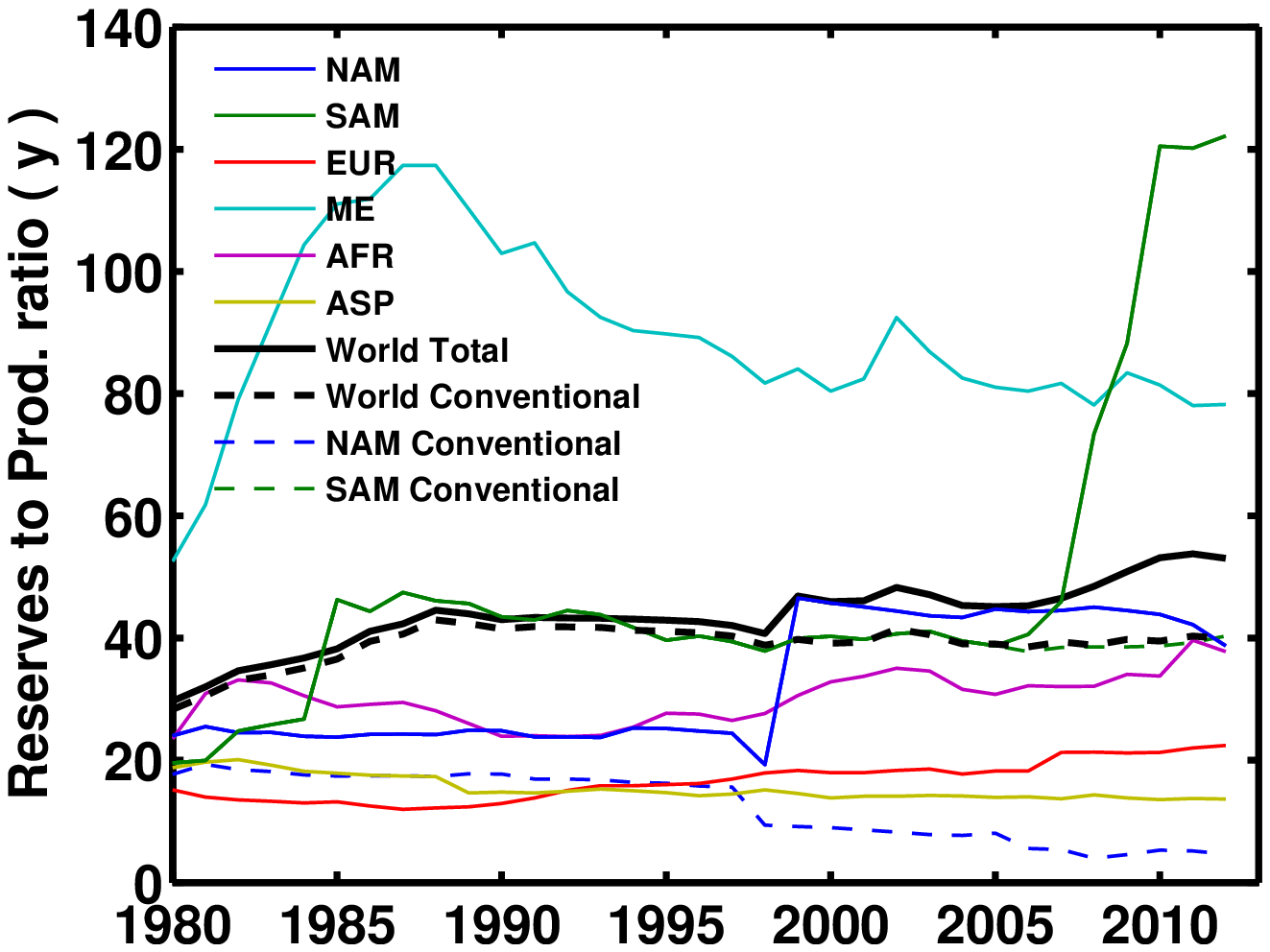}
		\end{center}
	\end{minipage}
	\hfill
	\begin{minipage}[t]{1\columnwidth}
		\begin{center}
			\includegraphics[width=1\columnwidth]{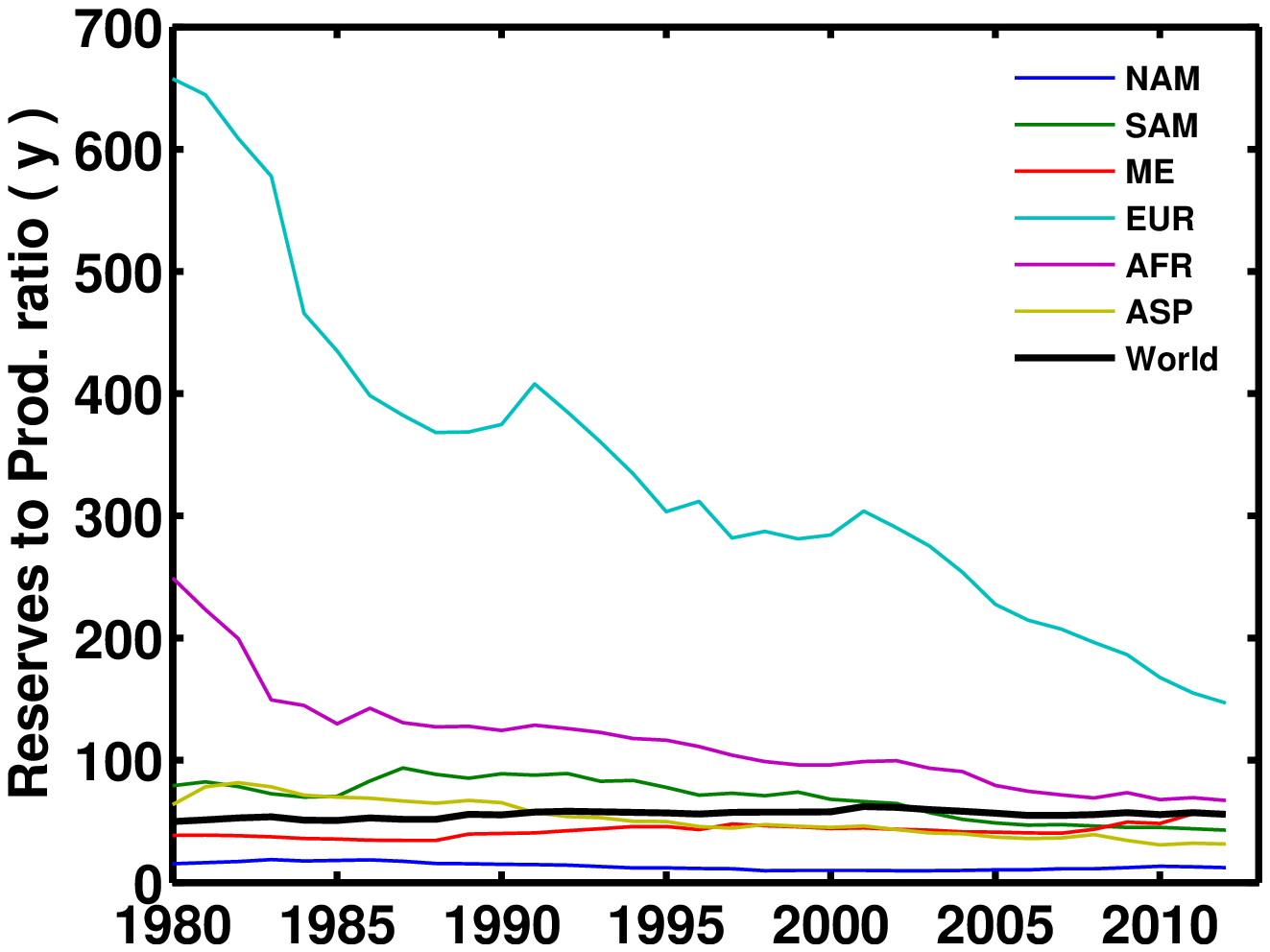}
		\end{center}
	\end{minipage}	
	\caption{Reserve to production ratios for stock resources for oil (\emph{left}) and gas (\emph{right}). This ratio is used to estimate to the inverse of $\nu_0$, in years. In the case of oil, this converges globally towards 44$\pm$10~years, while for gas it is of 54$\pm$6~years. The legend abbreviations correspond to North America (NAM), South America (SAM), Europe (EUR), Middle East (ME), Africa (AFR) and Asia-Pacific (ASP). The colour patches in the left panel correspond to amounts of reserves of unconventional oil, which were revised in \cite{BP2013} retrospectively, changing the results compared to \cite{BP2009}.}
	\label{fig:Production_to_Reserve}
\end{figure*}

The symbol $\nu_0$ represents the rate at which global reserves are consumed, or alternatively, the proportionality factor between increases in the size of reserves given increases in production, or vice-versa. It is a constant of the system that reflects \emph{both} the rates at which resources can be extracted or at which resource owners are willing to put them on the global market. $\nu_0$ can be evaluated from data using the ratio of global historical production to global reserves, i.e. what was known to exist at economic extraction costs in past years. This is shown in figure~\ref{fig:Production_to_Reserve} for oil and gas, for which the inverse, reserve to production ratios (R/P) were calculated from data from the BP Statistical Review of World Energy Workbook \citep{BP2009,BP2013}. Reserves continuously expand, but are continuously consumed as well \citep[for a discussion see][]{McKelvey1972}. The size of global oil and gas reserves has never been constant in history, neither has global production. However,  the ratios of production to reserves, when (and only when) taken at the global level, have been approximately constant in recent history, indicating that $\nu_0$ may well be simply a constant of time. The data is discussed in detail in the supplementary information supplied with this paper.\endnote{Note that the 2009 and 2013 BP statistical workbooks \emph{differ} in the case of oil, where historical data have been reclassified and unconventional oil resources were added to \emph{historical} reserves \emph{retrospectively} (tar sands and heavy oil in North and South America) and thus, BP data is not entirely reliable but only indicative. The R/P ratios differ, where it was constant at 42$\pm$2~years in the 2009 version and is not in the 2013 version, where it increases to up to 54~years in 2011. However, when one removes unconventional oil from the Americas, the ratio becomes constant again at 41 years with the 2013 version. This current large fluctuation in American reserves is likely to be related to the very large amounts of unconventional resources that can be reclassified as reserves above particular uncertain oil price thresholds in the vicinity of current oil prices, and thus this large amount is itself very uncertain and subject to future revisions. This does not occur in the case of gas. See the supplementary information for an extended discussion.} 

The behaviour of equation~\ref{eq:ResDep} is controlled by the \emph{a priori} unknown parameter $\nu_0$, which we estimated using BP data. On the regional level, various non-constant values for $\nu_0$ are observed (see fig.~\ref{fig:Production_to_Reserve}). Different regions have different energy policies related to their own political and geophysical situations. However, their aggregated output supplies the international demand for resources, which sets the price. This price moves up and down in order to adjust the upper value of cost that enables resources to be extracted; it defines the size of reserves out of the resource base. A large amount of trade occurs between regions of the world, and overall the value of $\nu_0$, on the global level, has been historically approximately constant. This supports the fact that the perceived price level, excluding short term fluctuations from speculation and hoarding except for a cost associated with risk due to volatility, evolves such that the ratio between production and reserves remains approximately constant, by \emph{unlocking} just the right amount of resources to include into reserves in order to supply demand, given that reserves are exploited at a rate of $\nu_0$.\endnote{$\nu_0$ can be seen as the time it would take to consume reserves at the current consumption level, and therefore has units of inverse years.} 

In the case of oil, (left panel of fig.~\ref{fig:Production_to_Reserve}), the R/P ratio converges towards 44$\pm$10~years, stabilising between 1987 and 2006, before fluctuating after 2006 when North and South American unconventional resources began to be considered. Its evolution before 1987 is related to the oil shocks of the late 1970s, where OPEC was formed and Middle Eastern production decreased faster than reserves, cartel formation forcing an intentional increase in the R/P ratio in that region. Its recent fluctuation after 2006 however is related to BP's inclusion of an uncertain part of the large amounts of both South and North American unconventional resources that are now considered to have extraction costs just below the economic threshold \citep[note that this includes significant government subsidies, which are furthermore evolving, fostering these fluctuations, see for instance][]{Greenpeace2010,IISD2010,IEAWEO2010}. In the case of gas, (right panel of fig.~\ref{fig:Production_to_Reserve}), this ratio is constant at 56$\pm$6~years throughout the period. In the case of coal, BP's global historical R/P ratio sees a declining trend since 1991 ending at 109 in 2011; however reserve data between sources do not agree and the economics of many coal resources appears poorly reported, while very large amounts are known to be in place \citep{WEC2010, BGR2010, MercureSalas2012}. We thus take a value for $\nu_0^{-1}$ of 125$\pm$50. For uranium, no historical reserve data is given by BP, but a similar process is assumed to take place, and for which the 2013 value for $\nu_0^{-1}$ is 16$\pm$1~years \citep[Data for uranium are taken from][]{MercureSalas2012, IAEA2009}. 

Given these uncertainty ranges, a sensitivity analysis is given in the supplementary information to explore the impact of changing the value of $\nu_0$ within these bounds, for oil and gas. It is shown that the uncertainty over $\nu_0^{-1}$, of $\pm 10$~years for oil and $\pm 6$~years for gas, contribute only a minor component of the total uncertainty, for instance much smaller than that associated with the uncertainty over the resource endowment, as shown in section~\ref{sect:Prices}. Adding a time dependence to $\nu_0$ within these bounds does not change these results, and thus it is not excluded that $\nu_0$ could change gradually but this has little impact. 

A constant value for $\nu_0$ is not an unexpected phenomenon. BP reserve data corresponds to a perception by the industry of the current outlook of energy resources, and their expectations regarding prices and global demand. It is natural to expect energy firms to increasingly expand their own conception of the economical limit of reserves by considering to develop new projects that were considered prohibitively expensive in the past, in order to \emph{maintain} reserves to a certain level, and that this level should evolve following global demand. This includes for instance the current arctic oil exploration frenzy, canadian tar sands, ultra-deep offshore rigs, etc. Fluctuations in reserve data are however also expected and known to arise, in particular when considering the controversy between using either so-called \emph{1P} and \emph{2P} reserve data \citep{Bentley2007}.

\section{Global overview of stock energy resources}

\subsection{Assumptions\label{sect:Assumptions}}

\begin{figure*}[t]
	\begin{minipage}[t]{1\columnwidth}
		\begin{center}
			\includegraphics[width=1\columnwidth]{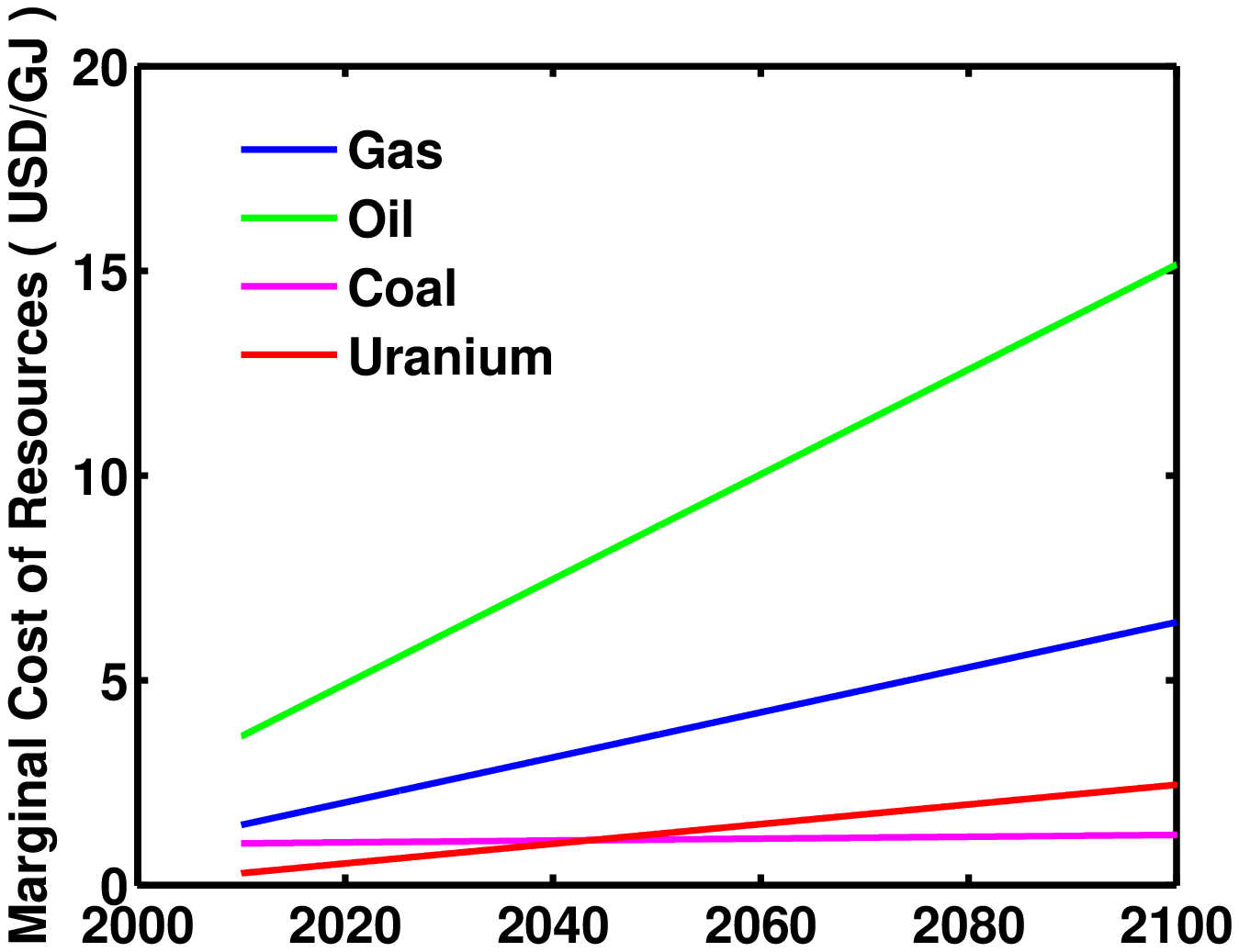}
		\end{center}
	\end{minipage}
	\hfill
	\begin{minipage}[t]{1\columnwidth}
		\begin{center}
			\includegraphics[width=1\columnwidth]{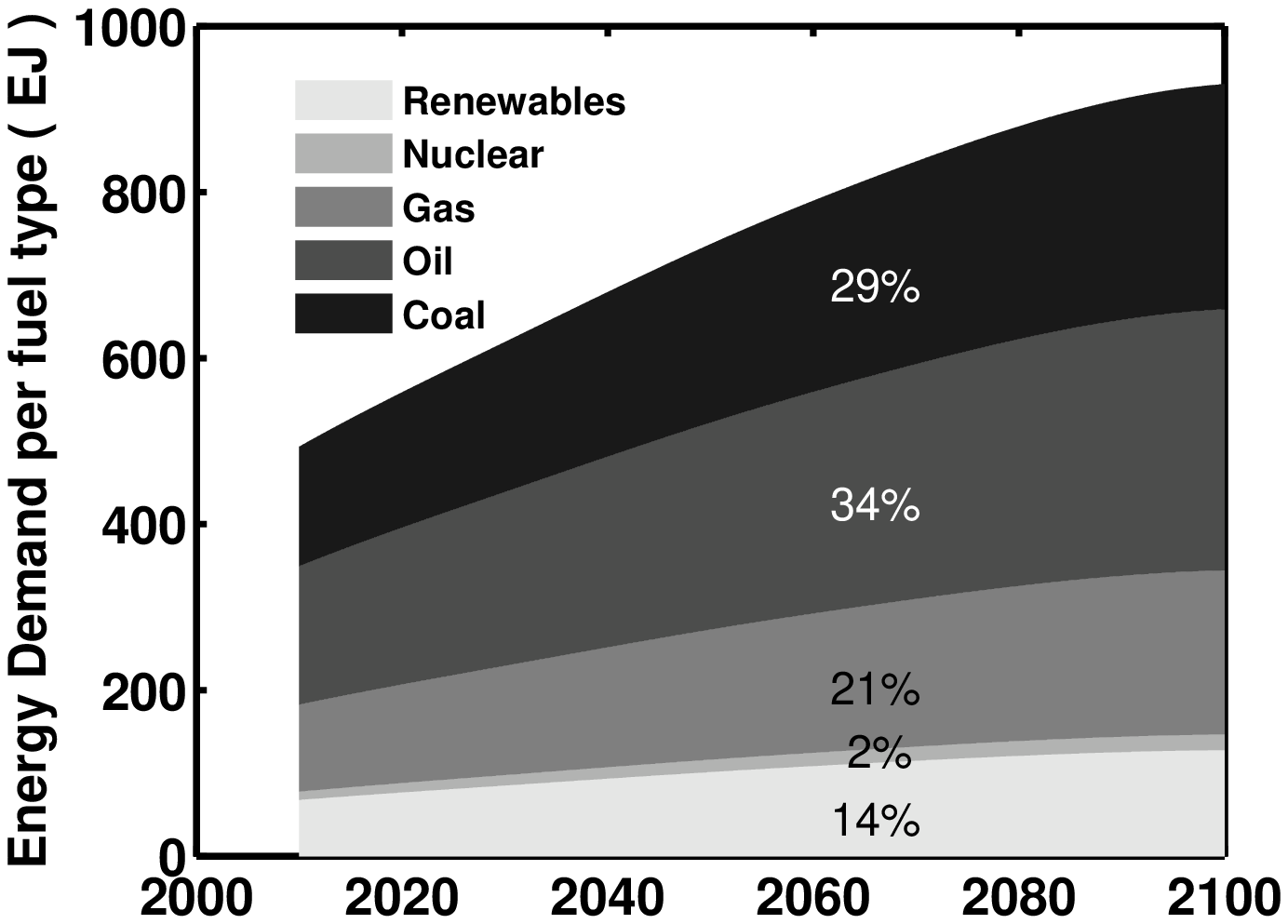}
		\end{center}
	\end{minipage}	
	\caption{Assumptions for calculations of stock resource flows and prices used in sections \ref{sect:Flows} and \ref{sect:Prices}. \emph{left} Exogenous linear marginal cost component of price assumptions used to calculate flows (\ref{sect:Flows}). \emph{right} Exogenous energy demand components used to calculate marginal cost components of prices (\ref{sect:Prices}). Note that here, the nuclear contribution to the total primary energy demand (TPED) appears as 2\% instead as the 6\% quoted in IEA TPED data \cite{IEAWEO2010} (IEA data for nuclear electricity production is 10~EJ). This is due to IEA's use of an arbitrary efficiency factor for converting 10~EJ of nuclear electricity into 30~EJ of primary nuclear fuel, out of 514~EJ of TPED. The value used here is 10~EJ out of 494~EJ of TPED, since the conversion factor of nuclear reactors is $already$ included in the resource data of \cite{MercureSalas2012}.}
	\label{fig:Assumptions}
\end{figure*}

In this section, two extreme modelling exercises are carried out to explore the model properties before providing an example of projection produced within a broader modelling framework. The assumptions for these exercises and for the projections are given here. The exploration of possible stock resource flows and associated world markets can be performed using either prices as exogenous in order to calculate resulting flows, or using flows (energy demand) as exogenous in order to calculate commodity prices, excluding in both cases the effects of hoarding and short term demand fluctuations. Values used or produced here and henceforth in this work correspond to \emph{marginal costs of production}, rather than real \emph{prices}, and whatever margins of profit, fixed costs and other cost component as well as fluctuations may be added to these marginal costs in order to construct real endogenous prices (prices calculated within an energy model as opposed to assumed), using separate assumptions, not done here, but left to the discretion of the modeller.

As a first exercise, we present in section \ref{sect:Flows} a calculation of stock resource flows given assumptions for the marginal costs deemed economic given energy carrier prices, denoted $P$. The assumptions are given in the left panel of figure~\ref{fig:Assumptions}. Values for $\nu_0$ are given in the preceding section. Given these, starting marginal costs required for supplying current demand were evaluated by finding which value of $P$ generates a $dN/dt$ that equals the current demand. Following this, rates for the increase of the carrier prices were used that generate an increase in supply consistent with current total primary energy demand increase, but were maintained constant throughout the century (linear prices). The resulting flows are given.

As a second exercise, we present in section \ref{sect:Prices} a calculation of the required marginal costs of energy production for each stock resource type for a scenario where total primary energy demand increases to around 900~EJ/y, but where the current shares of energy demand for these resources (coal, oil, gas, uranium) are rigidly maintained until 2100 (i.e. the structure of the current energy system is maintained). This energy demand curve until 2100 is within a range consistent with many recent projections, a review of which is given by \cite{Edenhofer2010}. The results are given in section~\ref{sect:Prices}. These rigid assumptions specifically exclude technology substitutions in both sets of calculations, an aspect which is explored separately in section~\ref{sect:RealSystems}.

Neither of these calculations produce realistic scenarios of energy use and prices, as we demonstrate in section~\ref{sect:RealSystems}. This is due to the absence of technology substitutions that enable to avoid price escalations related to scarcity by switching away from these sources, and of a dynamic feedback with energy demand from the economy. While the second aspect cannot be analysed here without a full blown description of extensive global macroeconomic modelling (e.g. using E3MG), the first is readily explored, however in the power sector only, by introducing the model described above into FTT:Power. FTT:Power enables to explore technology substitutions that are likely to arise in the power sector following changes in relative costs of energy systems, which include the cost of non-renewable fuels. Therefore, in the event of depletion and scarcity of some resources, the model endogenously generates switches of technology by gradually phasing out some types of systems, following possible rates of diffusion of new technology and rates of decommission of old systems \cite[see ][ for a complete description of the model]{Mercure2012}. Such changes reduce the demand for some types of resources, avoiding their depletion and large price increases that would occur in the case of a perfectly rigid demand. The interaction of FTT:Power with E3MG generates in general complex energy-economy interactions which are be explored elsewhere \citep{Mercure2013a}. Thus, for this work and for simplicity, two global energy demand scenarios were chosen based on IEA projections. The first is a baseline scenario of global energy policy grounded on the assumptions of the IEA's World Energy Outlook \emph{new policies scenario} \citep[see][appendix B, for a detailed description]{IEAWEO2010}. The second corresponds to a strong mitigation scenario aiming at reducing GHG emissions to below 50\% of the 1990 levels, following the assumptions of the \emph{450 scenario} of the World Energy Outlook, however with additional regulations and support for particular renewable technologies. Out of these scenarios, all exogenous inputs to FTT:Power were taken. This corresponds to the global demand for electricity and the global demand for energy resources that do not originate from the power sector (i.e. final use of coal, oil and gas by industries, households and transport). These assumptions are given in section~\ref{sect:RealSystems}. 

\subsection{Flows from stock energy sources for exogenous prices \label{sect:Flows}}

\begin{figure*}[t]
	\begin{minipage}[t]{1\columnwidth}
		\begin{center}
			\includegraphics[width=1\columnwidth]{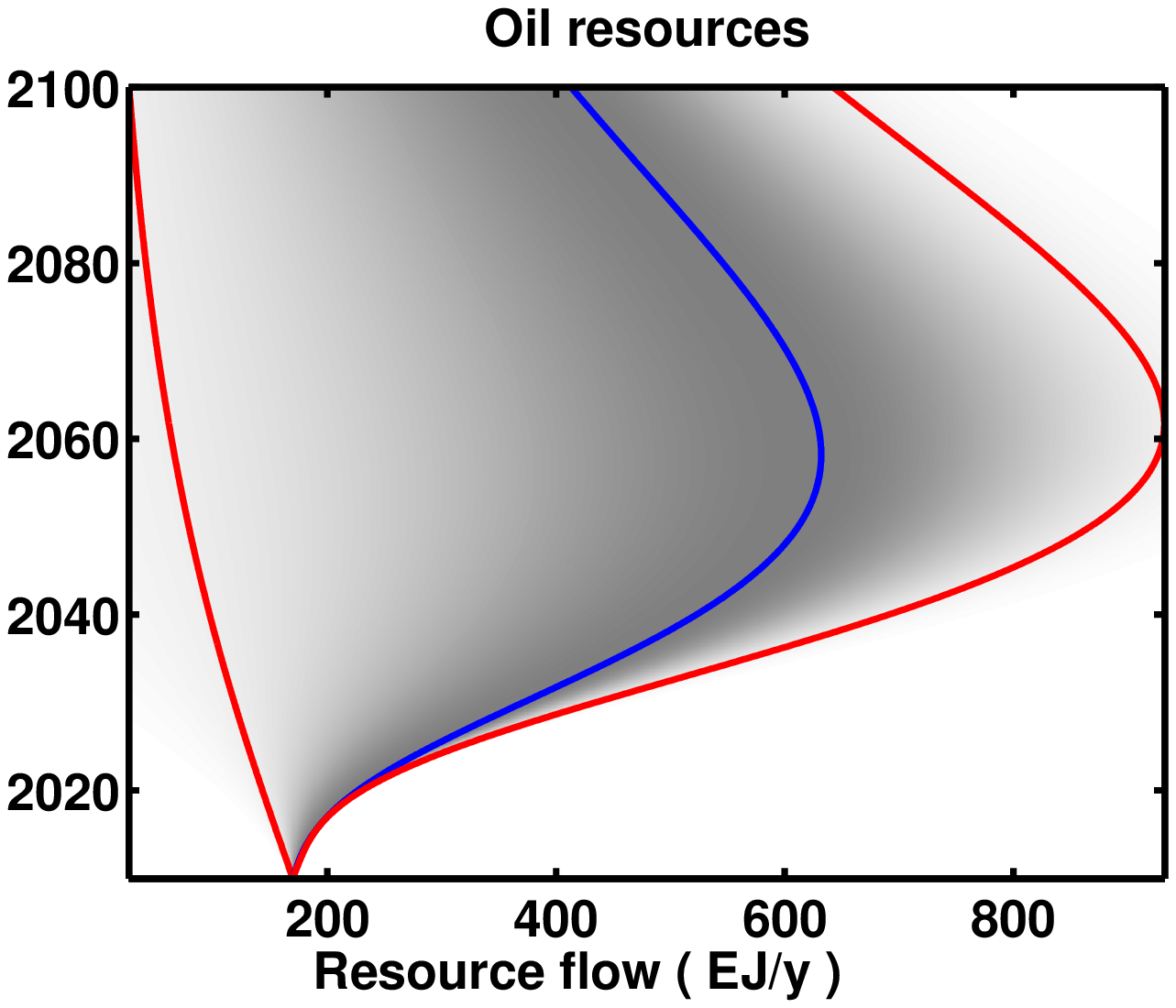}
			\includegraphics[width=1\columnwidth]{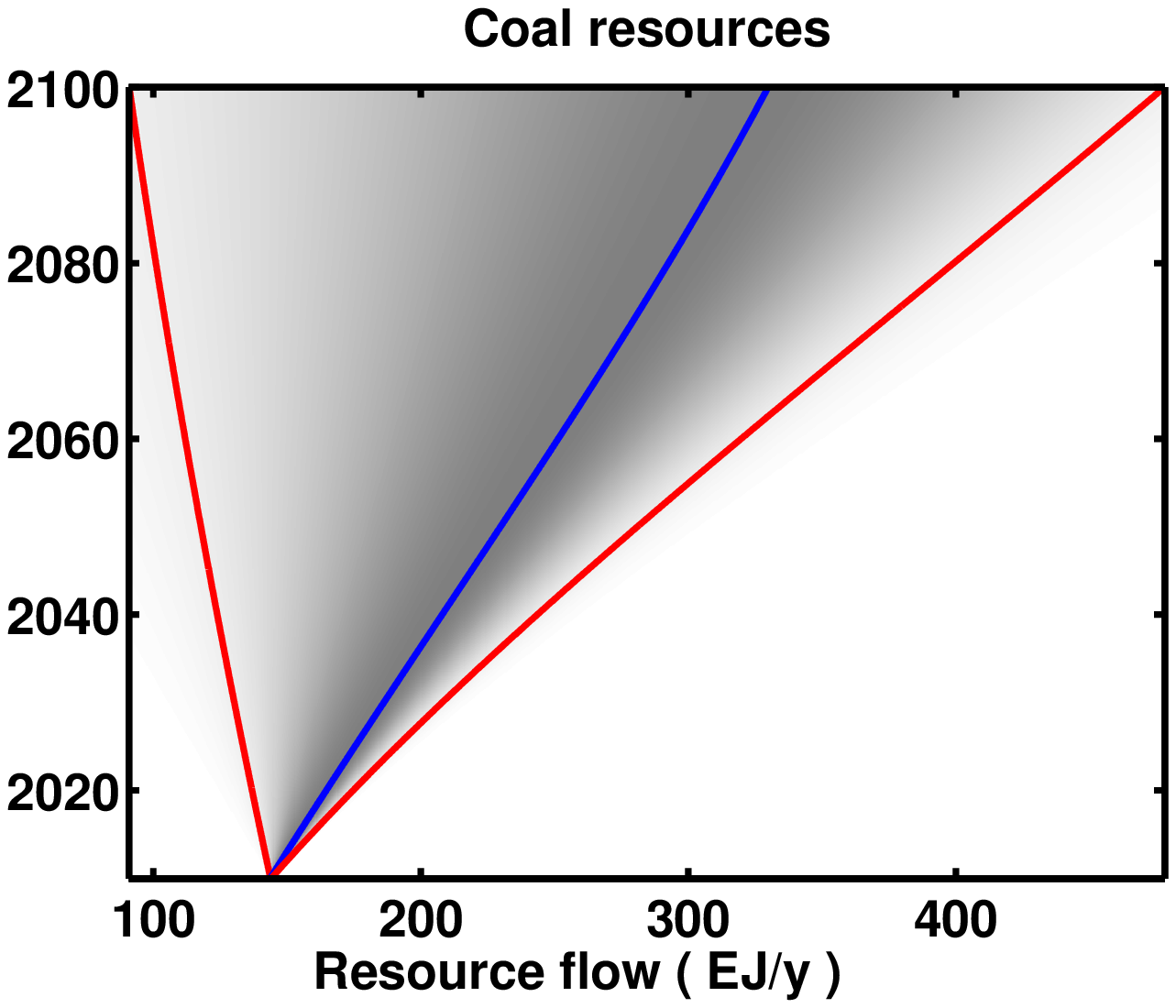}
		\end{center}
	\end{minipage}
	\hfill
	\begin{minipage}[t]{1\columnwidth}
		\begin{center}
			\includegraphics[width=1\columnwidth]{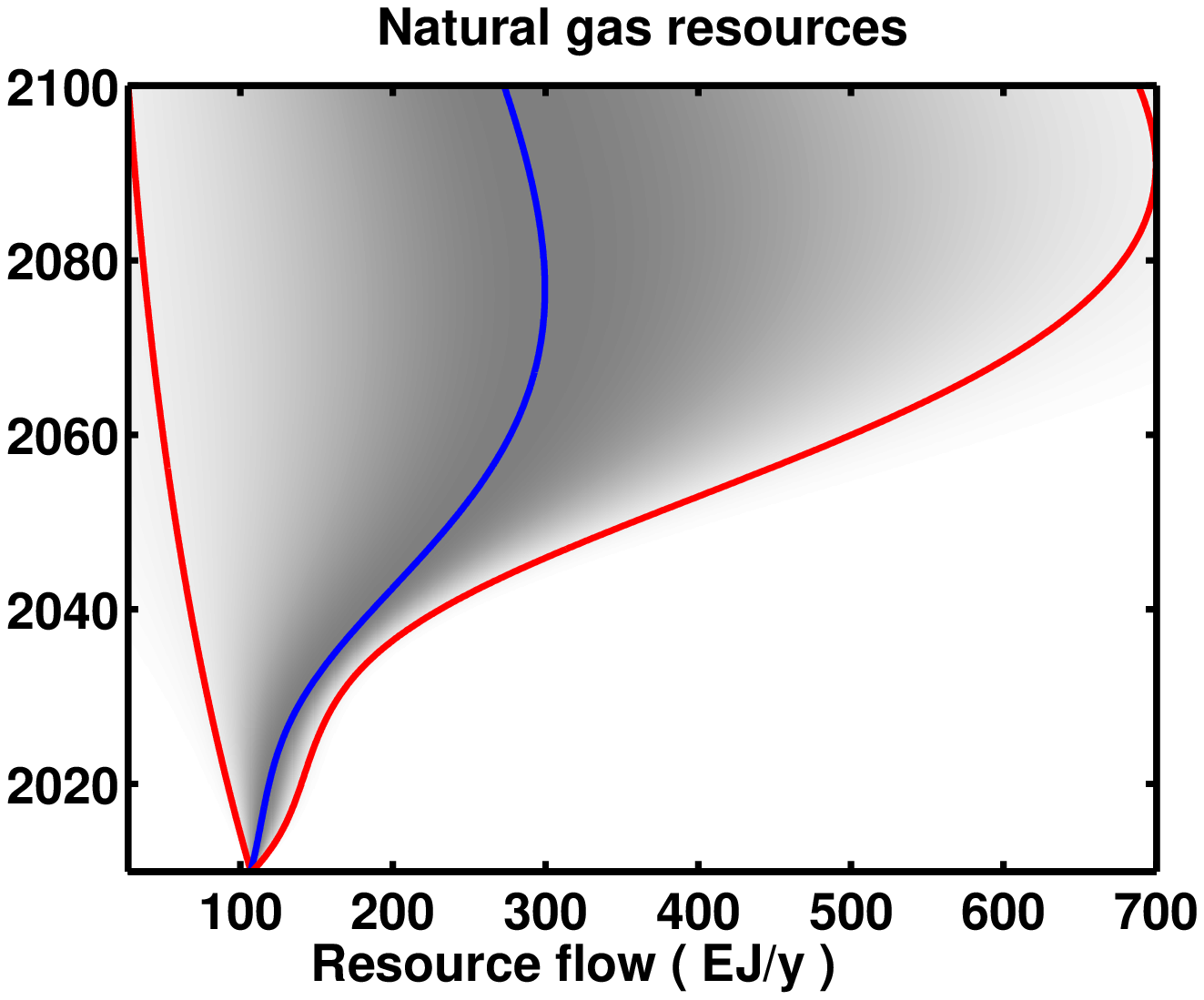}
			\includegraphics[width=1\columnwidth]{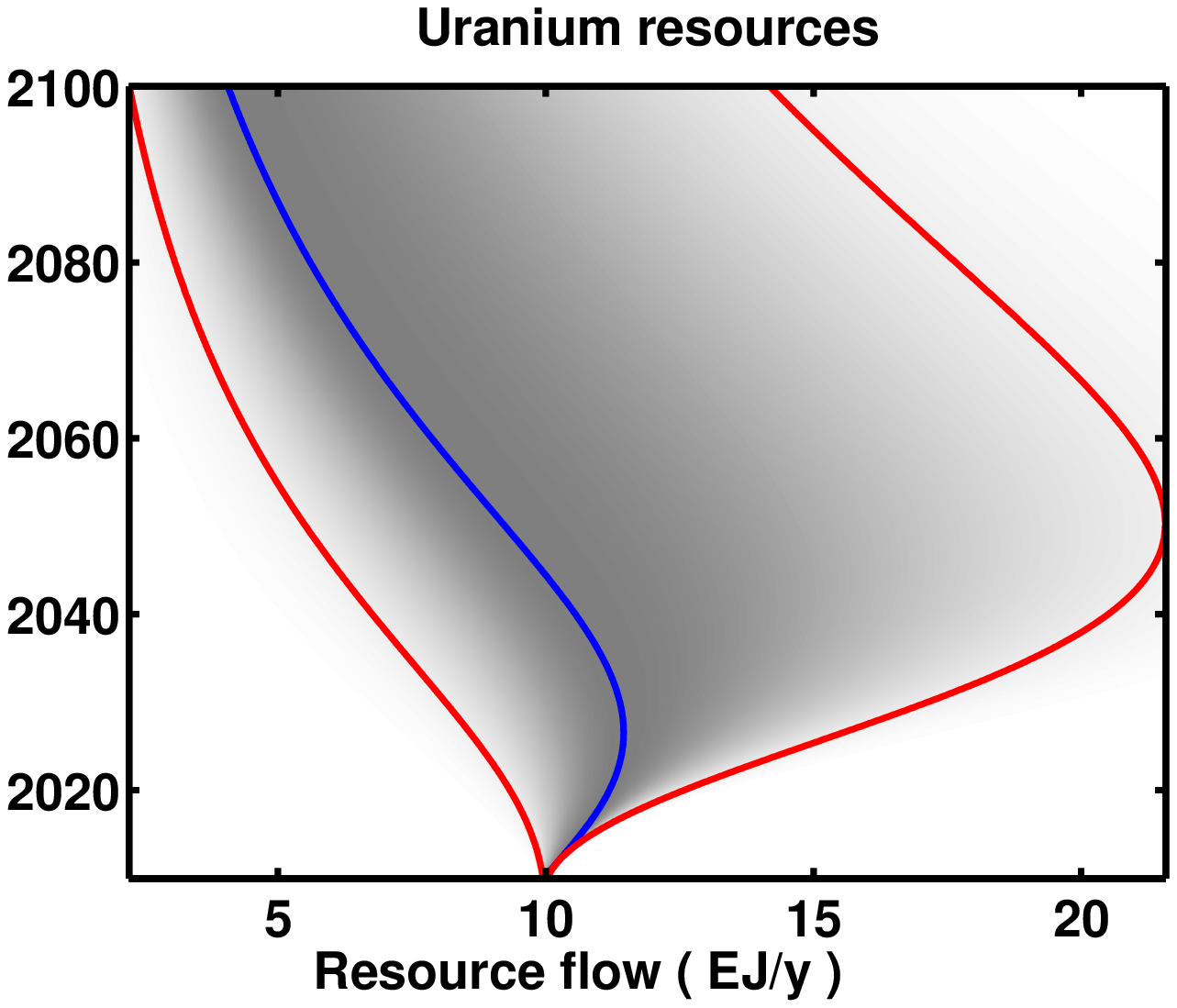}
		\end{center}
	\end{minipage}	
	\caption{Flows of oil, natural gas, coal and uranium, as calculated using equation~\ref{eq:ResDep} and the cost-supply curves given in figure~4 in \cite{MercureSalas2012}. The 96\% confidence level region is situated between the red curves, while the blue curves indicate the most probable flow values. The curves start in 2008 at current energy consumption values, given by the IEA.}
	\label{fig:GlobalStockFlow}
\end{figure*}

In the first extreme modelling exercise, scenarios for flows of stock resources were produced for oil, gas, coal and uranium using the resource distributions underlying the cost-supply curves given in figure~4 of our previous work \citep{MercureSalas2012}, current rates of resource exploitation (values for $\nu_0$ given above) and linear extrapolations of energy carrier marginal costs of production. In the case of coal, the sum of the distributions for hard and soft coal was used. The results are given in figure~\ref{fig:GlobalStockFlow}. In all cases, the curves start in 2008 at the current global production values reported by the \cite{IEAWEO2010},\endnote{Note that the IEA's value for nuclear electricity production of 10~EJ was used, not its reported value of primary nuclear fuel of 30~EJ. Efficiency factors for thermal reactors have already been taken into consideration in the cost-supply curve for uranium.} with zero uncertainty (i.e. current reserves are known). As time progresses, the increasing uncertainty in resources assessments at higher levels of use produces ever larger ranges of possible resource production values, or ranges of possible consumption paths, delimited by the red curves. These however must ultimately converge back to low flow values when peaking and depletion occurs.

Different results are obtained depending on the size of the various stocks. In the case of oil, which includes all types of unconventional oil, a peak in production occurs at around 2060, after which depletion begins. A similar situation occurs with natural gas, which includes all types of unconventional gas and methane hydrates, peaking later near 2080. Coal resources, however, are very large and depletion does not occur within a foreseeable future. It can only do so very far outside the time horizon of 2100. Resources of natural uranium, as reported by the \cite{IAEA2009}, are found to become depleted rapidly within the current century after peaking before 2025.\endnote{this excludes the reuse of fissile material available in nuclear waste, which would probably become economical $before$ the complete depletion of natural uranium resources. After 2025, the amounts of available fissile material in the waste produced by previous use of natural U will be large, and the high costs of recycling nuclear waste will eventually be equalled by the increasing costs of mining U ore which will become ever more difficult to reach.}

Potential flow values vary highly between resource types. Projected flows from oil resources are the largest, up to 600~EJ/y, giving however a faster rate of depletion compared to coal and natural gas. This is due to the current large rate of extraction to resource ratio $\nu_0$. Their massive expansion occurring after 2020 is due to the large scale exploitation of unconventional oil such as the tar sands and oil shales. While natural gas resources are smaller than those of oil, their depletion is projected further into the future due to a lower extraction rate to resource ratio $\nu_0$. Their massive expansion after 2040 is related to large scale exploitation of unconventional sources such as shale gas. 

In the case of coal, the rate of extraction is similar to that of gas, but their reserves are much larger, projecting the depletion far beyond 2100. In the case of uranium, given the small resource base, the low burn-up rates of current thermal reactors and the high value for $\nu_0$, the expected flow is very small compared to those of fossil fuels, but the depletion is projected to occur rapidly within this century. This indicates that dramatically higher conversion efficiencies are necessary to extend the resource base beyond the end of this century, which can be achieved with fast breeder reactors and/or using the thorium fuel cycle \citep{NuclearNuttall, Bonche2005}.

Note that these flow paths are not forecasts in any way. They are possible scenarios of resource use, given known (and uncertain) resources bases and realisable extraction rates and price paths, even though the price of carriers is hardly likely to follow a linear trend. Nevertheless, for any resource extraction path, the cumulative flow up to complete depletion must be equal to the technical potential of the resource, a requirement that has been carefully verified in these calculations. Therefore, for higher rates of extraction and resulting higher resource flows, depletion must occur slightly sooner than depicted here, and conversely, for slower resource use, depletion may occur slightly later. 

\subsection{Price paths for exogenous flows \label{sect:Prices}}

\begin{figure*}[t]
	\begin{minipage}[t]{1\columnwidth}
		\begin{center}
			\includegraphics[width=1\columnwidth]{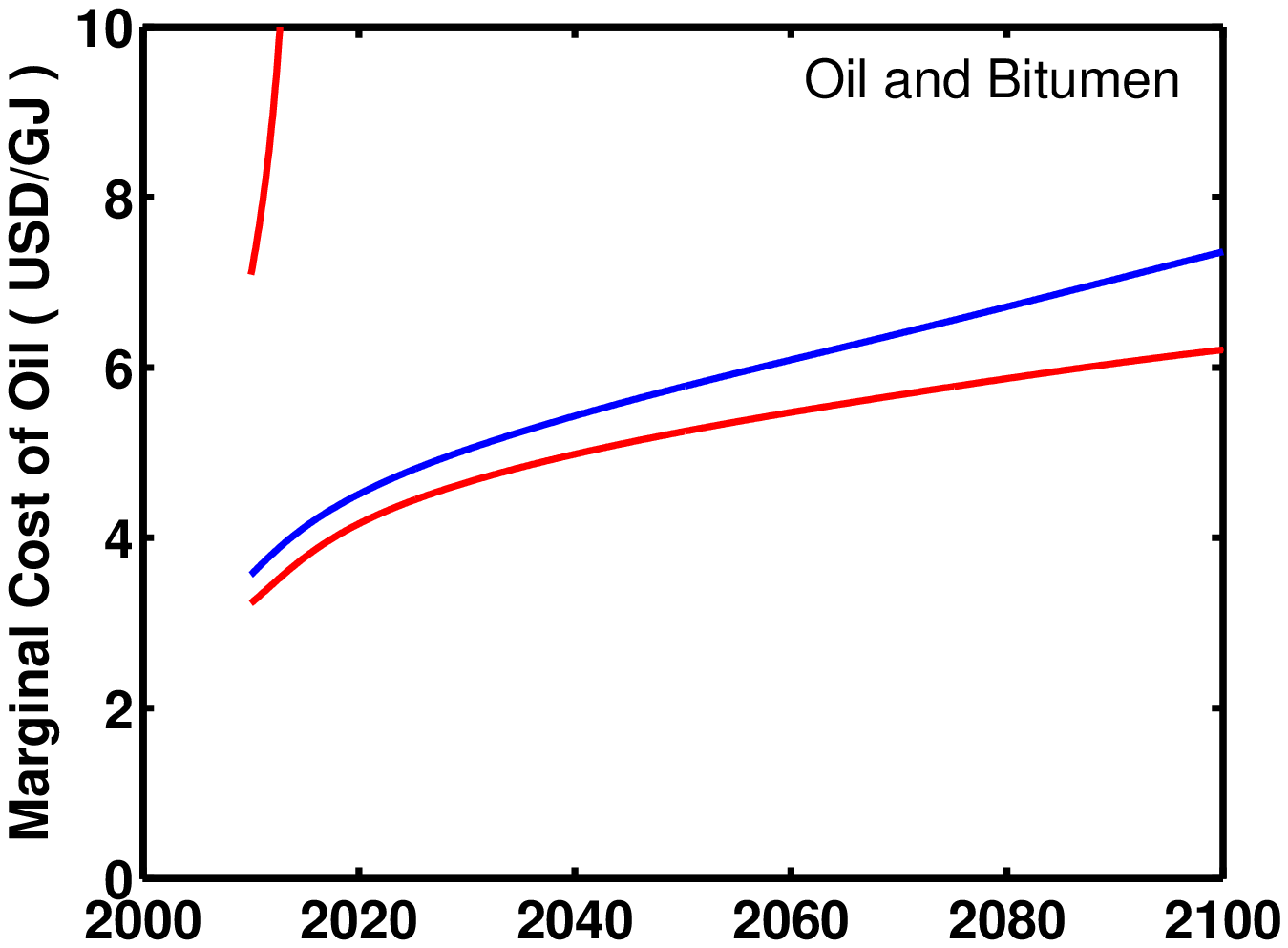}
			\includegraphics[width=1\columnwidth]{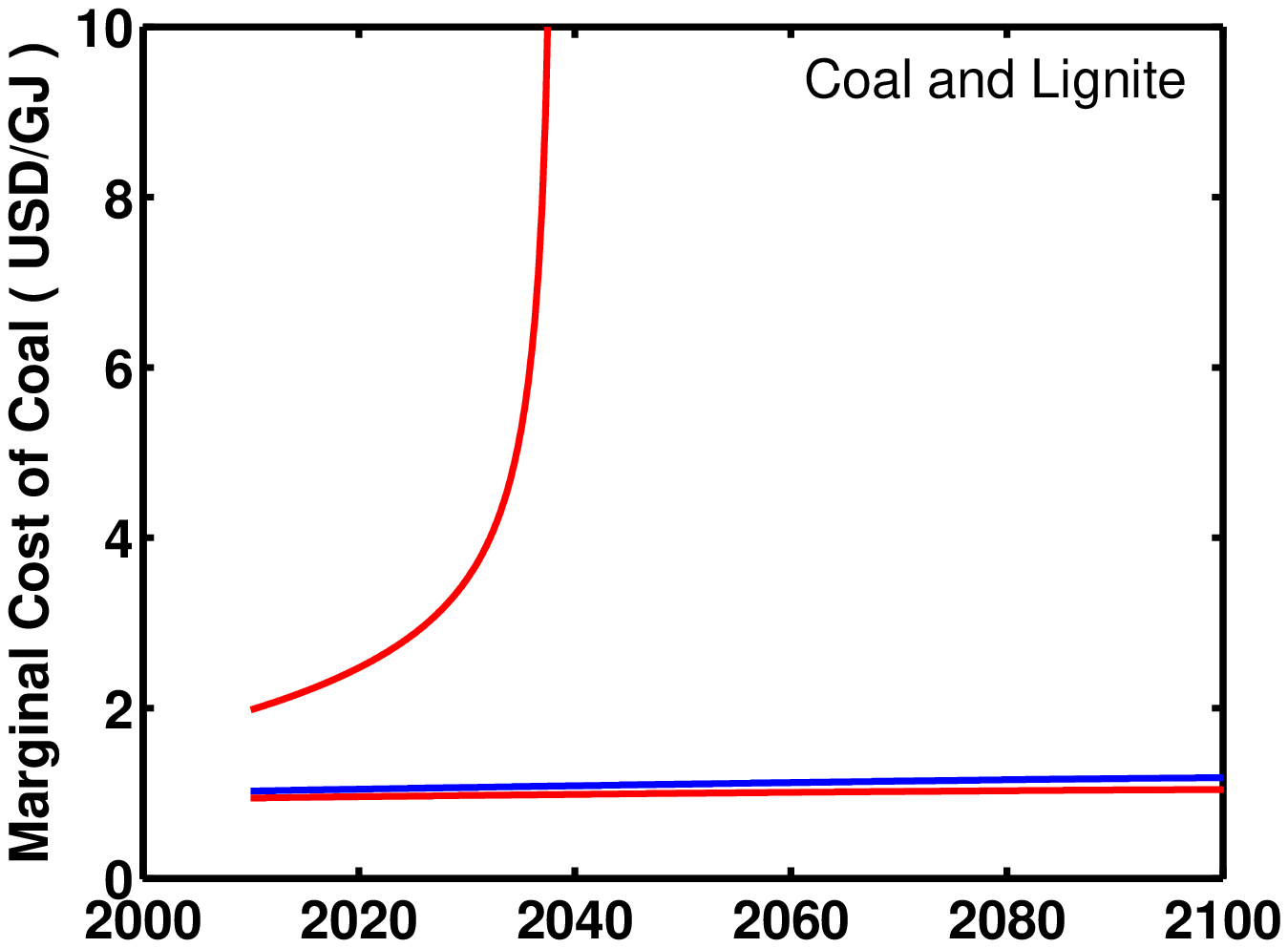}
		\end{center}
	\end{minipage}
	\hfill
	\begin{minipage}[t]{1\columnwidth}
		\begin{center}
			\includegraphics[width=1\columnwidth]{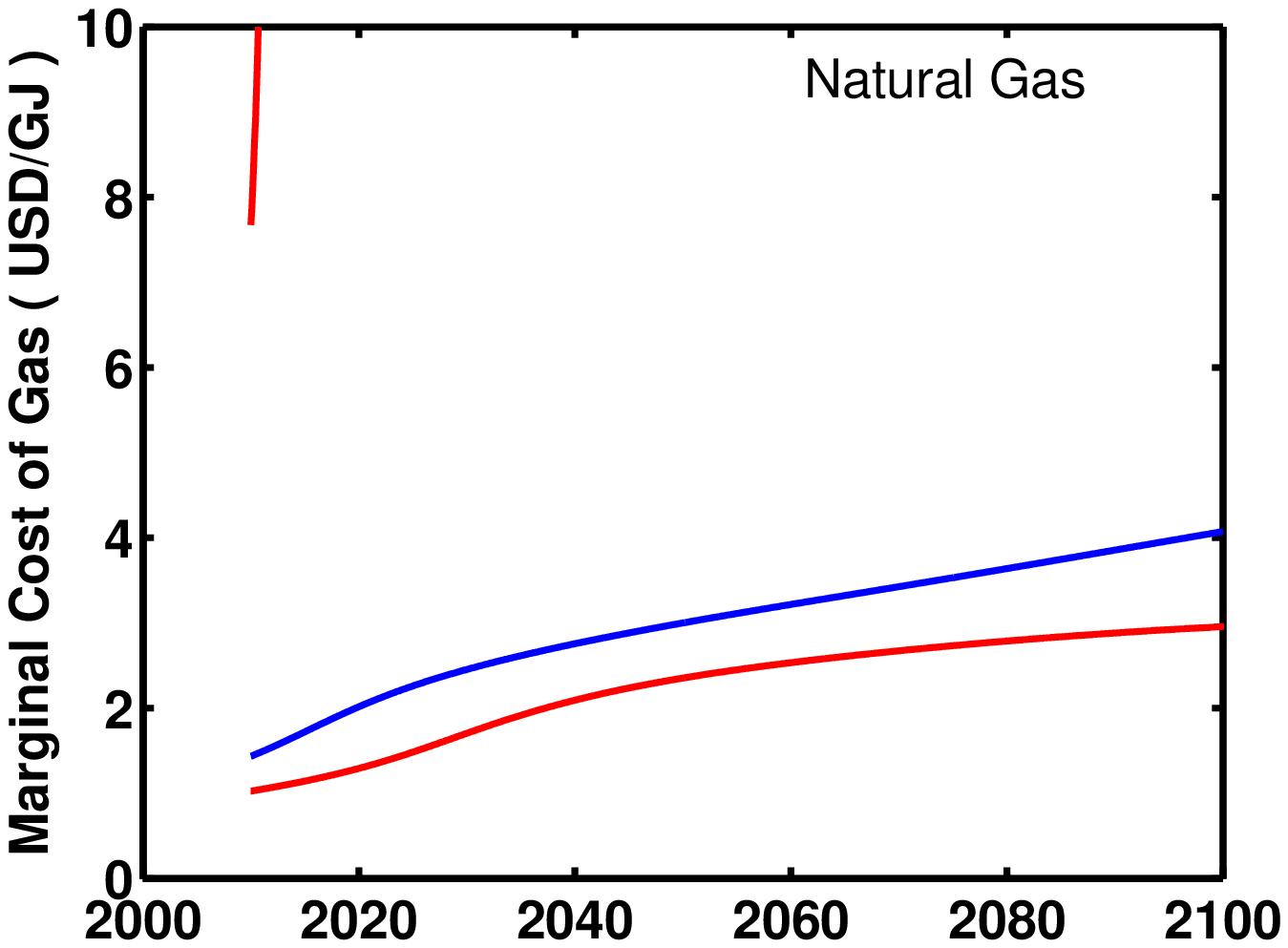}
			\includegraphics[width=1\columnwidth]{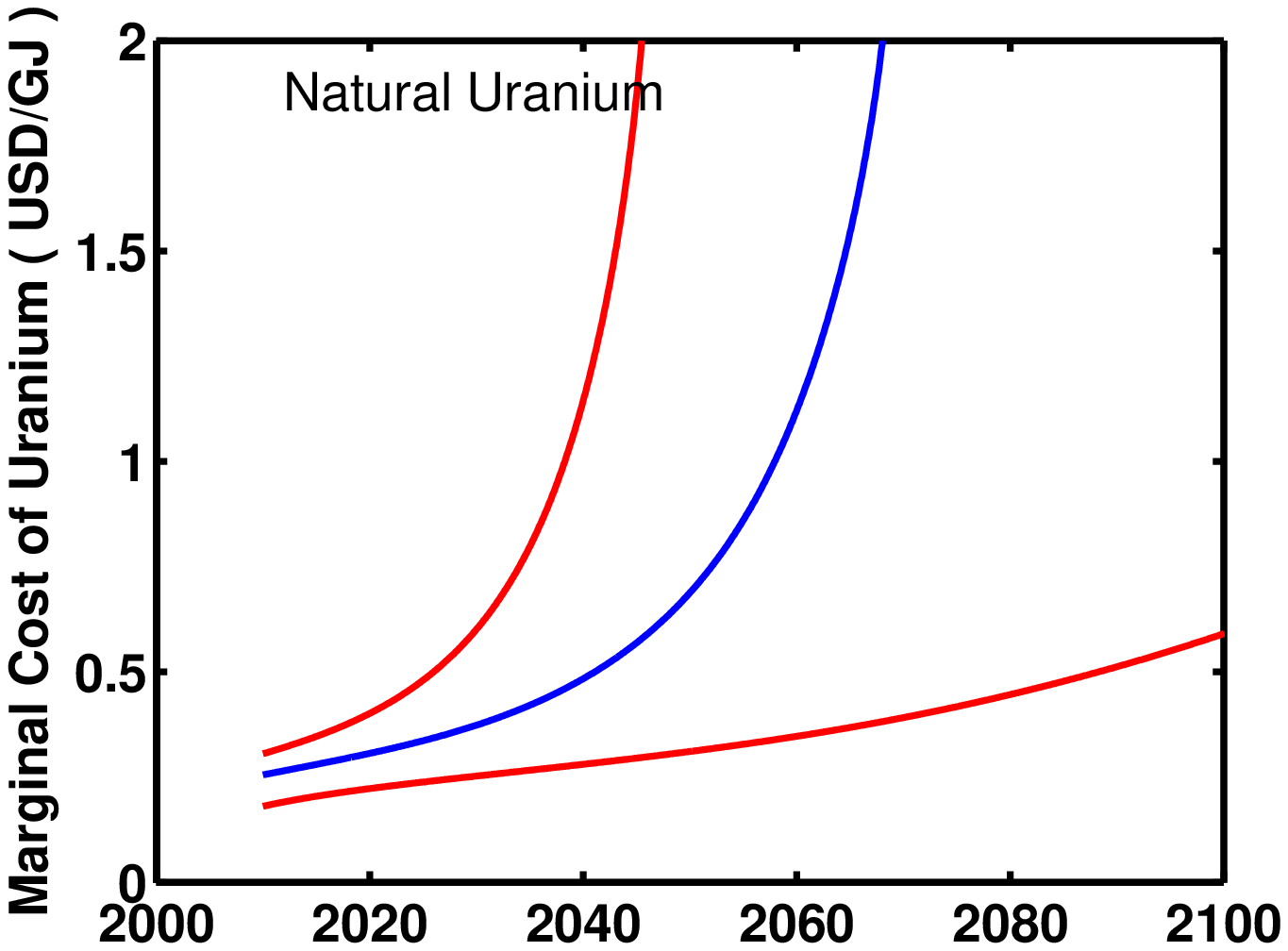}
		\end{center}
	\end{minipage}	
	\caption{Marginal cost calculations performed for four stock energy resources using the assumptions described in the right panel of fig.~\ref{fig:Assumptions}, which maintains the current composition of the energy sector, while expanding the total energy demand up to around 900~EJ, using the resource data from \cite{MercureSalas2012}. The blue curves correspond to the most probable values given resource uncertainty, while the red curves delimit the 96\% confidence range, where lower amounts lead to marginal costs increasing earlier.}
	\label{fig:InversePrices}
\end{figure*}

Conversely to the previous section, as a second extreme modelling exercise, the reverse problem is posed where one looks for the appropriate price of resources that unlocks just the right amount of resources to meet an exogenous (rigid\endnote{Rigid in the sense that the demand does not respond to prices.}) demand. In this case, the demand, or resource flow, is given as exogenous and the price, or marginal cost, is evaluated. This is done by performing an inverse calculation using a price optimisation of eq.~\ref{eq:ResDep}, such that the value of eq.~\ref{eq:IntResDep} is equal to the demand, separately for each type of energy carrier (oil, gas, coal, U). In such scenarios, it is possible that at a certain point in time, given the values of $\nu_0$, the remaining resource base cannot meet anymore the demand. In such a case, the price values gradually run away to infinity, signifying that all resources situated at all possible prices of extraction are under intense exploitation. Such a situation is very unlikely to occur, since the opportunity cost of using these expensive resources would become very large and other technologies and energy resources would be more cost effective, leading to technology substitution, switching away from that resource before it runs out. Alternatively, the global economy may also readjust its energy demand in order to avoid diverging prices of energy commodities. The divergence of prices therefore stems from rigid commodity demand values that do not respond to price signals. 

Figure~\ref{fig:InversePrices} presents the results of such an exercise, using the assumptions described in the right panel of fig.~\ref{fig:Assumptions}, where the current composition of the energy sector is maintained with a total energy demand scaling up to near 900~EJ/y in 2100. The blue curves correspond to the most probable resource bases given by the blue curves in fig.~4 of \cite{MercureSalas2012}. Meanwhile, the red curves delimit the 96\% confidence region, where the upper red curves correspond to the lower bounds for resources, while the lower red curves correspond to the higher end of the resource ranges. Therefore, in all cases, the marginal cost values calculated in the low end of the resource ranges diverge, while the curves for the upper ranges do not. 

In the case of oil and gas, gradual increases are observed in the marginal cost values, with a change in slope occurring between 2020 and 2030. This is related to the price enabling the accession to large unconventional resources, which include predominantly oil sands and shale gas respectively. The availability of these large resources tend to damp out possible future increases in price.\endnote{In a scenario where no additional environmental regulations prevent their exploitation, an obviously disputable assumption. In the event where such regulations arise (limiting fracking for instance, or  regulations being instated in Canada regarding river and land contamination from tar sands processing), the situation may become similar to the lower resource ranges given by the lower (high prices) red curves.} Meanwhile, in the case of coal, the marginal cost value is hardly affected by demand at all, unless resources turn out much smaller than expected, which is very unlikely. This is due to the very large resource being situated in a narrow range of extraction costs.\endnote{Note that for coal, it is not necessary to include low grade resources in high cost ranges, since normal grade resources are very large and unlikely to be consumed entirely within this century.} 

Finally, in the case of uranium, the marginal cost is expected to diverge, and the resource to run out, over the whole resource uncertainty range, if the current share of energy demand supplied by nuclear is maintained up to 2100, using current technology without recycling waste. At the current uranium burn-up rates, the resource base is insufficient. This indicates that the nuclear industry should either decrease its share of electricity generation significantly by 2100, or that much higher efficiency rates in resource used per unit of electricity produced are achieved before then, involving possibly a much higher rate of recycling of nuclear waste than occurs at present. Future energy systems planners will adopt either of these solutions in order to avoid this projected fuel cost escalation.

\subsection{Real systems: allowing technology substitutions \label{sect:RealSystems}}

\begin{figure*}[t]
		\begin{center}
			\includegraphics[angle = -90, width=1.5\columnwidth]{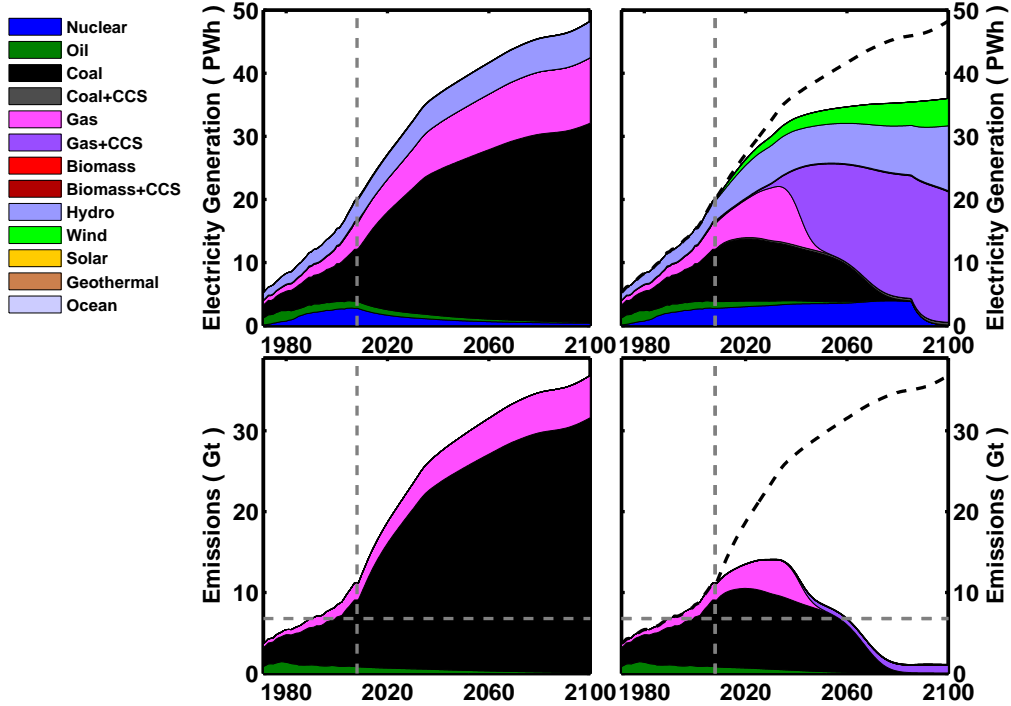}	
		\end{center}
	\caption{Example of changes in the power sector that could generate strong increases in the marginal cost of natural gas and uranium, in comparison to a baseline scenario, where in the \emph{top} panels electricity generation is given, while in the \emph{bottom} panels the associated emissions are shown. On the \emph{left} panels is shown the baseline scenario, which consists primarily in assuming that current policies extend into the future. On the \emph{right} is shown a mitigation scenario where carbon pricing exists (starting at 22\$/tCO$_2$ and increasing by 2\% per year ) and support for the wind industry and CCS technology. Dashed black curves correspond to the same baseline as in the left panels, given for comparison. The 2008 technology emission factors were obtained by comparing emissions to electricity generation in IEA data, maintained in the projection. Horizontal dashed lines give reference to the 1990 levels, while vertical dashed separate projections from historical data.}
	\label{fig:GandE}
\end{figure*}
\begin{figure*}[t]
		\begin{center}
			\includegraphics[angle = -90, width=1.5\columnwidth]{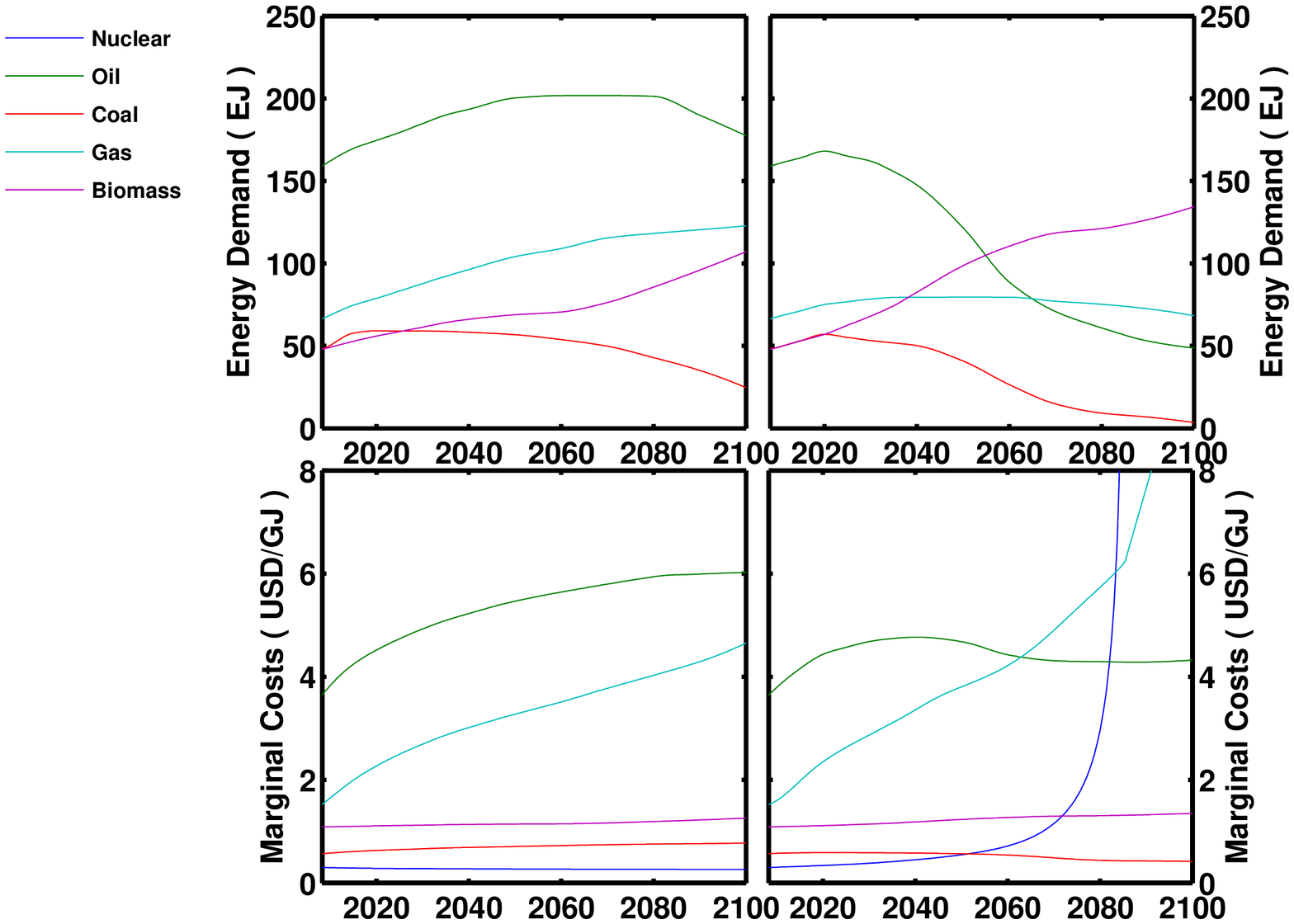}	
		\end{center}
	\caption{\emph{top} $panels$ exogenous assumptions for the demand for coal, gas, oil, uranium and biomass, that does not originate from the power sector, for the baseline (\emph{left}) and the mitigation (\emph{right}) scenarios given in figure \ref{fig:GandE}. \emph{bottom} $panels$ Marginal cost of production for coal, gas, oil and uranium that result from the global demand and the theory presented above, along with the marginal cost of biomass production which is obtained directly from the cost-supply curve.}
	\label{fig:Demand_Prices}
\end{figure*}

Although the modelling exercises given in the last two sections provide insight on the scale of available resources and on the process of their gradual consumption, they both depict limiting situations that are very unlikely to occur. This is due to the facts that: 
\begin{description}
\item 1- There exists a feedback between prices and demand in the global economy 
\item 2- Technology/resource substitution processes occur that enable reductions of the demand for specific commodities.
\end{description}
Prices are not likely to remain strictly linearly increasing as in section \ref{sect:Flows}, and the composition of the demand is not likely to remain fixed in the future, as in section \ref{sect:Prices}. Effectively, as depletion progresses, marginal costs of exploitation increase and prices increase, and these induce gradual technology switching and/or reductions in overall energy demand. Complete technology switching away from a particular fuel occurs when the price of this fuel makes its use uneconomical. Therefore, the prices can never escalate to very high values as long as technology switching options exist since switching away occurs before the price diverges at complete depletion, and thus stock resources are never depleted entirely. Technology switching is however constrained by capital lifetimes and can therefore take some time to take place, in particular in the power sector.

The inverse calculation problem described in section \ref{sect:FlowEqn} to derive a marginal cost of production given an exogenous demand was introduced into the model of the global power sector FTT:Power, which specifically simulates technology switching given plant lifetimes and dynamic rates of technology diffusion, described in length in \cite{Mercure2012}. It thus provides an appropriate testing ground for this theory, an exercise that also generates direct insight on the effect to prices of the future composition of the power system constrained by natural resources. Figures~\ref{fig:GandE}~and~\ref{fig:Demand_Prices} present the results of simulations performed with this version of FTT:Power for all four non-renewable resources, oil, coal, gas and uranium, and the values of $\nu_0$ evaluated above. The particular example given was chosen on the basis that it describes well the properties of this model in order to demonstrate its validity, not for the goal of recommending any particular power sector technology scenario. 

The theory presented here however requires the global demand from all sectors for natural resources, not entirely provided by FTT:Power, which treats the power sector only. Thus, the missing components of the demand unrelated to electricity production had to be taken as assumptions, described below.\endnote{The current work on building models of technology substitutions in the transport, industrial and household sectors will in time replace these fixed assumptions.} In all other respects, the model assumptions are very similar to those of the baseline presented in \cite{Mercure2012}.\endnote{The baseline here features no carbon pricing and no subsidies or taxes on any technology.} These simulations were performed for two sets of policy and demand assumptions, a baseline scenario and a mitigation scenario. The top panels of figure~\ref{fig:GandE} present electricity generation by type of technology for both scenarios, the baseline on the left and the mitigation scenario on the right. In these plots, the dashed vertical lines provide a visual reference to the present and the data to the left of these lines corresponds to historical data, while on the right are given the calculated projections. Meanwhile, in the bottom panels, the associated emissions from fuel combustion in the power sector are given, where the horizontal dashed line indicates the 1990 level. 

The demand for energy commodities (coal, oil, gas and uranium) was calculated using endogenous values for fuel demand by the power sector, and exogenous values for fuel demand from the rest of the economy. Demand values for the rest of the economy are given in the top panels of figure~\ref{fig:Demand_Prices} for each scenario. In the baseline, the demand for oil, originating primarily in the transport sector, was assumed to peak late in the century, motivated by a gradual transition to alternative transport technologies. The non power-related demand for gas, originating primarily from the industrial and buildings sectors, was assumed to increase gradually up to 2100, although slowing down due to gradually increasing overall efficiency in parallel with an increasing demand for heating services. The non-power demand for coal, originating primarily in the industrial sector, was assumed to rapidly peak and gradually decrease due to technology switching and increased use of natural gas. Finally, the demand for biomass was assumed to gradually increase, at a rate accelerating in the second half of the century due to a higher diffusion of biofuels for transport. In the mitigation scenario, oil demand peaks rapidly due to massive technology switching in the transport sector towards biofuel and electric cars. The demand for biomass increases sharply to supply this additional demand. The non-power demand for natural gas however remains relatively constant and the demand for coal sharply declines due to massive technology switching and electrification of industrial processes.

The bottom panels of figure~\ref{fig:Demand_Prices} present the resulting marginal cost of production of coal, gas, oil and uranium in these model runs of FTT:Power including the present theory for both scenarios, which can be used, with additional chosen assumptions, to construct endogenous prices. In the baseline scenario, a strong increase in the cost of natural gas is observed, associated to an increasing global demand, forcing the price to enable the extraction of shale gas and more expensive resources. This rate of increase is however damped past 2020-2030 due to the large amount of resource available at these cost ranges. Meanwhile, the marginal cost of coal hardly changes, irrespective of the sharply increasing demand, reflecting the sheer scale of low cost coal resources. Nuclear reactors are mostly decommissioned and see a sharp decline in the baseline scenario, leading to a decrease in the price of uranium. Finally, the price of oil increases due to increasing depletion, but the rate of increase is damped by the accession to massive amounts of unconventional oil. However, since the main component of the demand does not originate from the power sector but the exogenous transport demand, the analysis of oil demand is outside the scope of this work.\endnote{A similar treatment of technology substitution in the transport sector would be required in order to produce a dynamic demand for oil, electricity and biofuels that responds to prices, which will be the subject of future work and a new model, FTT:Transport.} 

In the mitigation scenario, strong support is given to wind energy through a subsidy (35\% of the LCOE throughout the simulation period), as well as through the pricing of CO$_2$ emissions (starting at 22~\$/t and increasing by 2\% per year up to a value of 140~\$/t in 2100), while moderate support is given to electricity production using capture and storage (CCS) technology (10\% of their respective LCOEs).\endnote{Additional subsidies are given to biomass based electricity and solar technologies, of 35\% and 50\% respectively, without much impact in this particular scenario. These are effectively pushed out of the market by wind and gas turbines.} The introduction of large amounts of variable renewable electricity into the grid requires increases in the amount of flexible type of generation, which can be provided for by, for instance, gas turbines, oil plants, hydroelectricity or energy storage.\endnote{Note however that the construction of storage systems that would make a significant difference for grid balancing is very large \citep{Mercure2012}, and additionally, the valuation of their benefits for electricity markets is not properly taken into account for large scale deployment but could be done in the future \citep{Zafirakis2013}.} This motivates a massive expansion of the gas turbine technology into the electricity market, which eventually dominates. The carbon pricing however motivates the installation of CCS on all gas turbines by 2060, reducing drastically emissions. Meanwhile, the pricing of carbon makes coal technologies gradually come out of favour, while the nuclear industry maintains a constant market share. However, with the assumption that nuclear reactors only use natural uranium, do not recycle waste and maintain the low conversion efficiency of ordinary thermal reactors, and that thorium or fast breeder reactors are not considered \citep[see][ for a discussion of the various nuclear options]{MercureSalas2012}, the resource base is seen unable to maintain the share of nuclear capacity and a strong price increase for natural uranium is observed, which generates a decline of the nuclear industry starting at around 2070.\endnote{Note that although the price of U goes off the scale of the lower right panel of figure~\ref{fig:Demand_Prices}, it does not diverge to infinity. Its scale of increase is related to the rate at which the nuclear industry can decommission its power stations in a scenario where waste recycling is not allowed.} The massive expansion of gas turbines generates a stronger increase in the cost of gas compared to the baseline, which is however damped slightly due to the large amounts of shale gas available. The cost of coal resources gradually decreases following the decline of coal in electricity generation. With a smaller demand for oil resources by the transport sector as assumed exogenously, the price of oil initially increases but stabilises and decreases slightly in the middle of the century.

Finally, global power sector emissions are given for both scenarios in the bottom panels of figure~\ref{fig:GandE}. While emissions increase monotonically up to 37~Gt/y in 2100 in the baseline scenario, they peak at 14~Gt/y in the mitigation scenario in around 2030, they decrease afterwards monotonically and reach the 1990 level of 7~Gt in 2060, and then decrease to a low level of 1~Gt in 2080 where it remains constant. Thus the policy assumptions for the mitigation scenario are not stringent enough to reach 50\% reductions by 2050. Note that the technology mix outcome of the various possible subsidy schemes transform this situation greatly. While this example is not particularly attractive for policy, it was chosen because it displays strong marginal cost changes related to changes in demand, where for instance, the development of biomass based electricity was pushed out of the market by the strong mutually beneficial combination of wind and gas turbines, which is not necessarily beneficial for overall emissions.\endnote{For instance, negative emissions from biomass gasification with carbon capture and storage can reduce emissions by much larger amounts.} Detailed scenario analyses using FTT:Power are not the primary objective of this work and will be presented elsewhere. This scenario does, however, warn of the potentially strong effect that energy policy and future demand can have on the price of energy commodities.

This section demonstrates that the combination of the theory presented above for treating dynamically the consumption of stock resources, with a model of technology substitution, enables to effectively project future marginal costs of energy commodities given exogenous demand values from the rest of the economy. However, since technology substitutions can occur in all sectors of the economy, additional flexibility in demand values exists in the global economy. Therefore, while the power sector amounts to a very significant share of global fuel use, more accurate calculations for energy commodity prices can be performed using a combination of this theory with a complete family of models of technology substitution for all major fuel users of the global economy, for instance the FTT family \citep[for the theoretical framework underlying the FTT family, see][]{Mercure2012b}, the development of which will enable to remove one by one the exogenous demand assumptions given above. This is a substantial project which is currently under way and will be the subject of forthcoming publications.

\subsection{Impacts on climate policy}

Carefully designed climate policy must take into consideration the amounts of low emission energy resources available in the world in order to produce its desired outcomes. Additionally to this, however, climate policy making must evaluate its own effects on energy prices, the effects of energy prices on technology substitution and onto the economy.

Four aspects must be considered in such policy frameworks: 
\begin{description}
\item 1- The cost distributions of energy resources,
\item 2- The relationship between energy prices, the size of the resource bases and rates of exploitation,
\item 3- The damping effect of technology substitution on these prices,
\item 4- The effects of climate policy onto energy prices and their subsequent consequences onto the well-being of the global economy, including through the price of electricity.
\end{description}
The theory presented above, in combination with our previous work \citep{Mercure2012, Mercure2012b, MercureSalas2012} is appropriate to treat these issues, for instance by using the model FTT:Power, excluding the effects of high energy prices on the economy, which can be modelled with an economic model such as E3MG. 

Failing to address these issues, by for instance being optimistic on the amounts of low carbon resources available, is likely to lead to badly planned energy policy where high energy carrier prices result (e.g. the price of uranium) and/or strong rebound effects arise due to reductions in prices, generating additional demand for these carriers if the price difference is not absorbed by energy pricing policy (e.g. energy taxes), where for instance reducing the consumption of oil in one region of the world reduces the oil price, and induces additional consumption elsewhere. Optimistic reports on the availability of low carbon resources abound in the literature, including the World Energy Assessment \citep{WEA2001} and the IPCC special report on renewables \citep{IPCCSRREN2011}, which list energy economic potentials as single values rather than cost distributions, making these considerations ambiguous. Additionally, reports exist that claim the feasibility of the complete replacement of fossil fuels by renewables by 2050 including a report by the World Wildlife Fund \citep{WWFEcofys2011}, which do not take consideration of either cost distributions of resources, provide reliable estimates of energy resources or treat the feasibility of massive rapid diffusion of new technology into the marketplace. Such overly optimistic studies must be taken with care by the policy community. Energy systems are complex and deserve to be treated using complex dynamic modelling. 

\section{Conclusion}

We have presented a model that, along with our previous paper \citep{MercureSalas2012}, completes our work on the economic potentials of energy resources by generating marginal cost of non-renewable energy carrier production, which can be used to project future energy prices. We have defined a model, based on cost distributions of non-renewable energy resources published in previous work, that simulates the simultaneous consumption and expansion of energy reserves. This model uses a dynamic differential equation approach to calculate at each time step of a projection the amounts of energy resources consumed at a particular price. This generates an endogenous resource consumption path given an exogenous price path. This model can however be used in reverse, where the price that generates an appropriate resource supply is found by optimisation, generating an endogenous price calculation given an exogenous demand. Using both approaches, ranges of possible consumption paths as well as possible marginal cost ranges have been derived given the available resource bases and associated uncertainty ranges. The use of rigid demand or price assumptions however lead to unrealistic price escalations or depletion timescales. By connecting this model to one of technology substitution based on rules of investor choice over pairwise comparisons, price increases are damped through substitution, avoiding the depletion of particular resources and the associated price increases. The analysis presented thus generates insight for energy planning related to non-renewable resources. It generates consumption peaking timescales for oil, natural gas, coal and uranium resources, where in particular, emphasis is given to the limited amounts of available natural uranium. When enabling technological change, technology substitutions are observed that avoid marginal cost increases. Reflections on the policy implications of the model and results presented are provided.

This model forms the third part of a theoretical framework for exploring global future technology transformations and climate policy, which consists in (1) a technology substitution model based on a coupled family of logistic differential equations representing technology diffusion into the marketplace, (2) a database and theoretical basis for tracking the use of global energy resources, and (3) this work, which provides the basis for a model for endogenous price formation. This cornerstone will enable the completion of the full FTT family of technology models which aims at projecting multi-sectoral global greenhouse gas emissions.

\section*{Acknowledgements}
The authors would like to acknowledge T. S. Barker for guidance, D. J. Crawford-Brown for support as well as H. Pollitt, P. Summerton and U. Chewpreecha at Cambridge Econometrics for highly informative discussions. We are furthermore grateful to an anonymous referee whose remarks led us to improve the paper significantly. This work was supported by the Three Guineas Trust (P. Salas and J.-F, Mercure),  Conicyt (Comisi\'on Nacional de Investigaci\'on Cient\'ifica y Tecnol\'ogica, Gobierno de Chile) and the Ministerio de Energ\'ia, Gobierno de Chile (P. Salas), and the UK Engineering and Physical Sciences Research Council, fellowship number EP/K007254/1 (J.-F. Mercure).

\section*{References}

\bibliographystyle{elsarticle-harv}
\bibliography{../../../CamRefs.bib}

\theendnotes

\newpage
\onecolumn
\setcounter{section}{0}
\renewcommand*\thesection{S.\arabic{section}}
\renewcommand*\thesubsection{S.\arabic{section}.\arabic{subsection}}
\part*{Supplementary Material}

This documents presents supplementary information to the paper entitled `On the global economic potentials and marginal costs of non-renewable resources and the price of energy commodities', \emph{Energy Policy}. While this additional material would reduce the clarity of the main paper with unnecessary detail, it is essential for anyone wishing to explore further the assumptions of the model, the data justifying its theoretical structure and the mathematical properties of the equations. Thus, this document presents, in order, an exploration of BP reserve data in their 2009 and 2013 versions, the differences observed and how these are used to determine the parameter $\nu_0$ of the model (section~\ref{sect:v0}), a sensitivity analysis of the model over the value of $\nu_0$ for oil and gas consumption and price paths (section~\ref{sect:sensitivity}) and an exploration of the mathematical properties of the model, in particular hysteresis and path dependence (section~\ref{sect:math}). 

\section{Discussion of Reserve to Production ratios and the determination of $\nu_0$\label{sect:v0}}

\subsection{Reserve data and its reliability}

Reserve data is subject to continuous revisions and changes due not only to energy commodity price changes in time but also from political motives. Such data is therefore in general difficult to use and interpret. For the model presented in this paper, it is however necessary to use reserve data since a statement is made on the dynamics of reserves and their relation to commodity prices. The data used, taken from the BP Statistical Review of World Energy Workbook, was explored using two different versions of the workbook, those of 2009 and 2013. In the case of oil, differences were observed between the two versions of the BP data, over historical trends.\footnote{Thanks to comments from the anonymous referee, this might have otherwise escaped our attention.} Therefore, it is apparent that BP changed its historical oil reserve data \emph{retrospectively}. Our interpretation of historical reserve data concerns what we consider that the oil industry thought the reserves were each year in the past. However, from this observation we know that BP data are subject to later revisions and thus this interpretation of the data is only partially reliable. No such significant changes were observed in the case of gas.

The details of the changes however, present in every region in almost every year, can be analysed and the most significant changes can be traced to two regions and particular cases: unconventional oil in NAM and SAM. These correspond to special categories in the BP workbook for Canadian tar sands and Venezuelan heavy oil, of which the reserves have been revised by very large amounts retrospectively (e.g. the 2013 workbook reports much larger tar sands reserves for 1999-2012 than the 2009 workbook). When removing these unconventional reserves from the data, however, the changes are much less significant. This is shown in figure~\ref{fig:Production_to_Reserve}. The top panels show total reserves in the 2009 workbook (left) and the 2013 workbook (right), while the bottom panels show the same data minus unconventional reserves and highlights the addition of unconventional oil . It is noted that without unconventional reserves, no significant data changes are apparent, and that large fluctuations appear in recent years for unconventional oil. The retrospective data change justifies to characterise these increases as fluctuation in this context.

Unconventional reserves are known to be on the edge of the profitability threshold, and may actually well set the marginal cost. In these high cost ranges, large amounts of unconventional resources exist, in other words, the cost-distributed density sees a very sharp rise at these cost values. Therefore, small variations in the estimated \emph{economic} extraction cost (or changes in the value of existing fossil fuel subsidies) may signify bringing in large amounts of resources into reserves, the actual real amount of unconventional oil that is economical to exploit being highly speculative. In addition to this, important (direct or indirect) subsidies exist in unconventional oil producing countries with the specific goal to help bring their exploitation cost within the economic threshold. All of this contributes to generate large fluctuations in the current amounts of oil reserves, and since production does not fluctuate in this way, the R/P ratio fluctuates along with reserves. 

\begin{figure}[t]
	\begin{minipage}[t]{.5\columnwidth}
		\begin{center}
			\includegraphics[width=1\columnwidth]{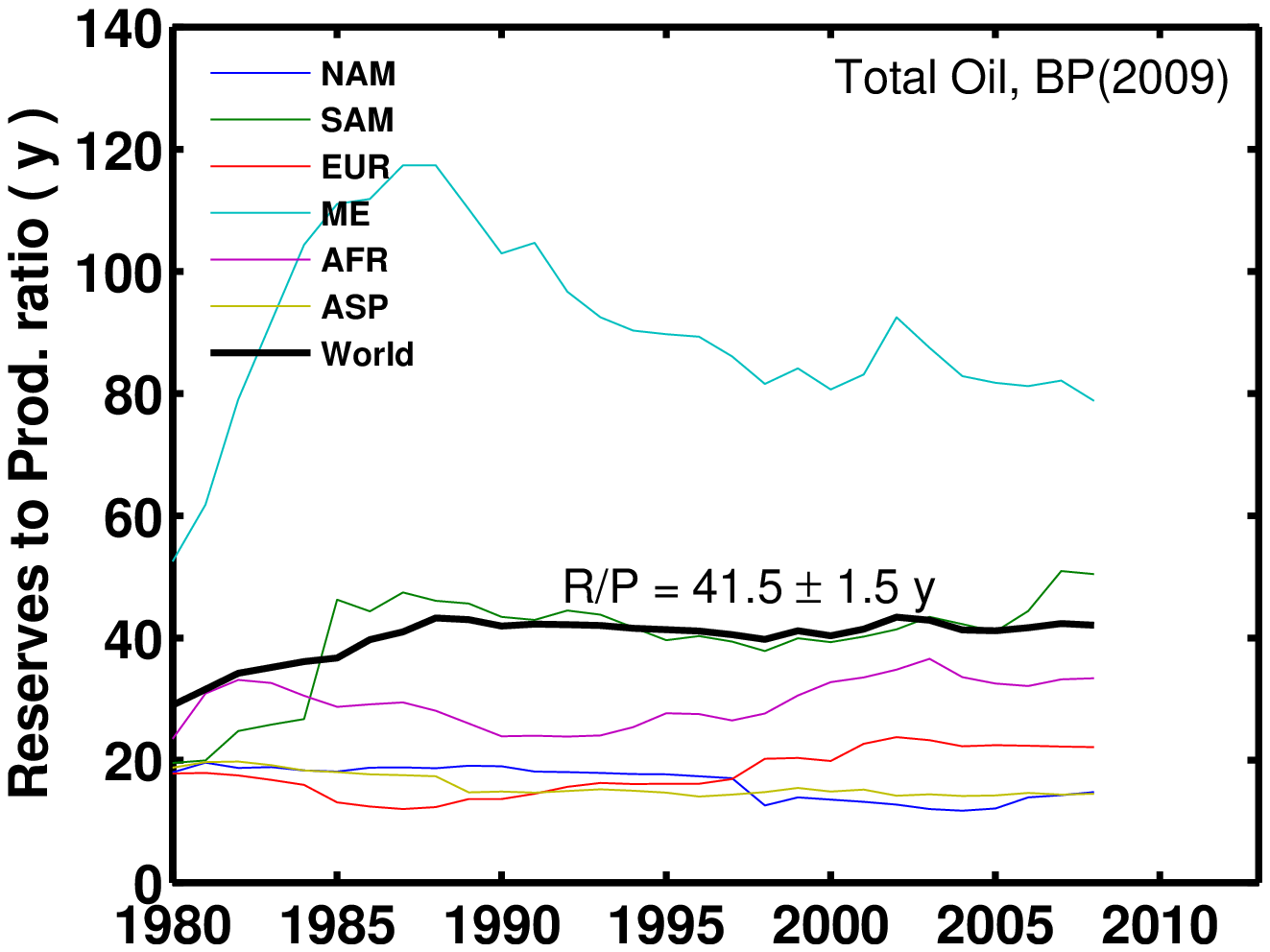}
			\includegraphics[width=1\columnwidth]{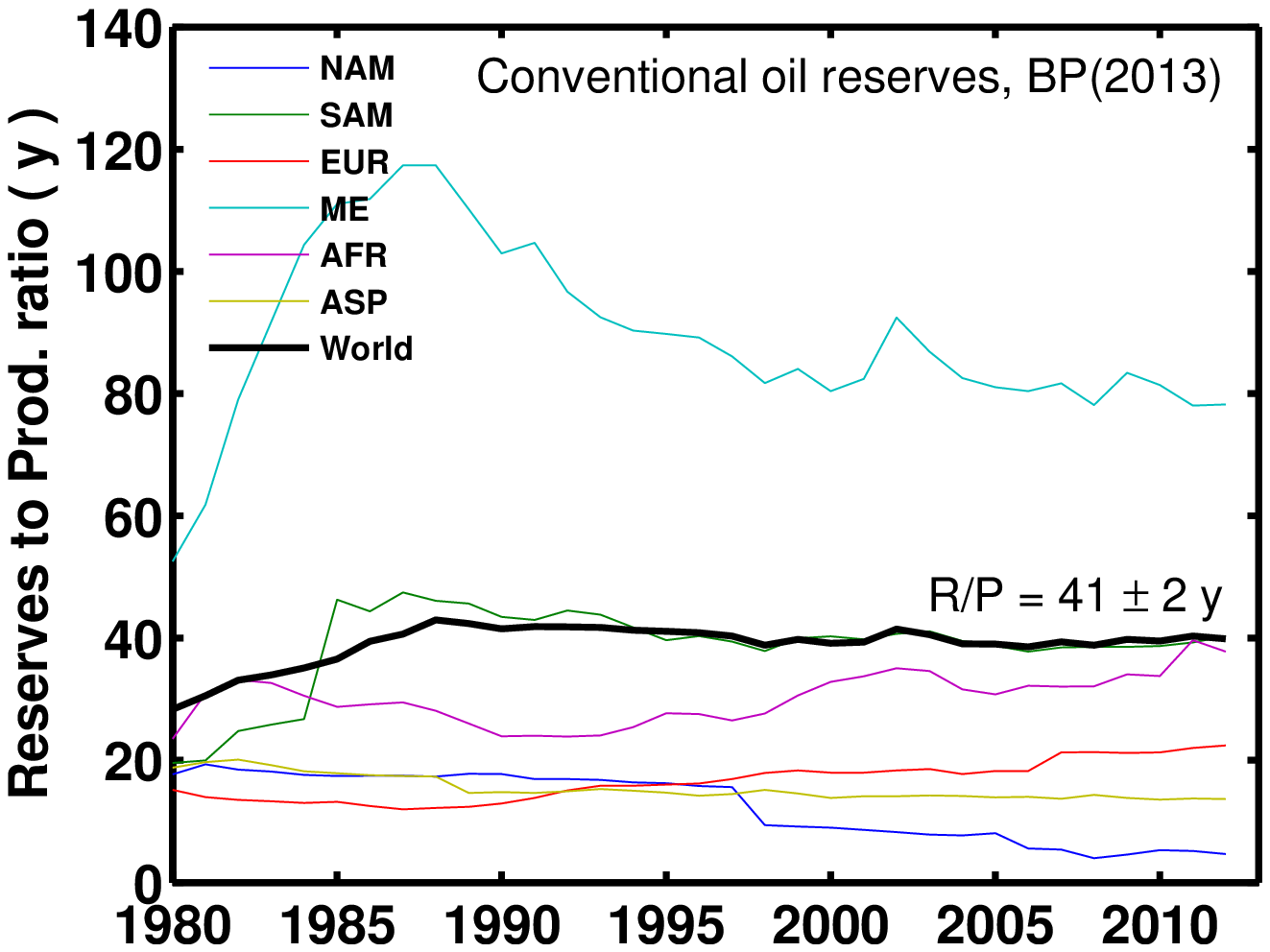}
		\end{center}
	\end{minipage}
	\hfill
	\begin{minipage}[t]{.5\columnwidth}
		\begin{center}
			\includegraphics[width=1\columnwidth]{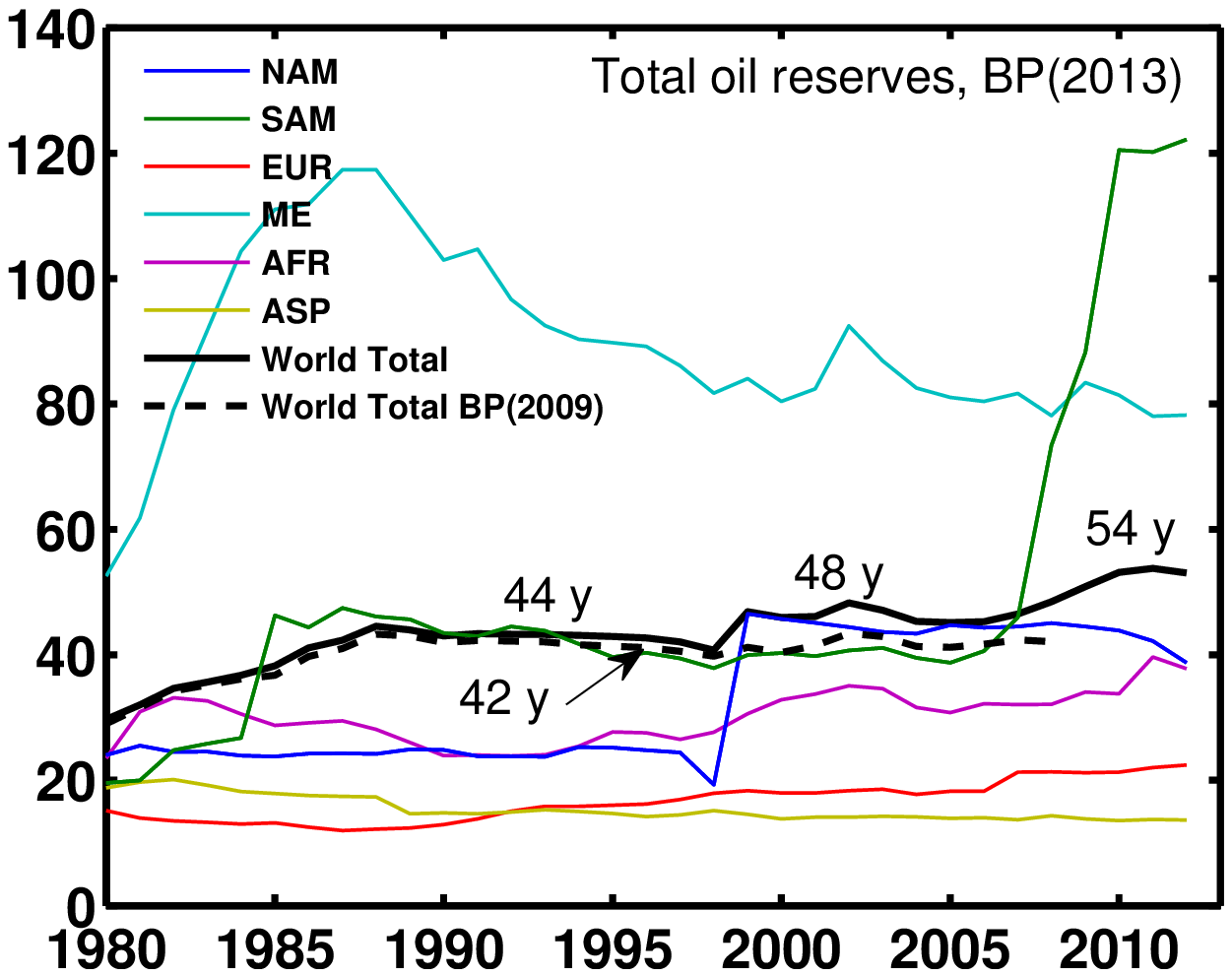}
			\includegraphics[width=1\columnwidth]{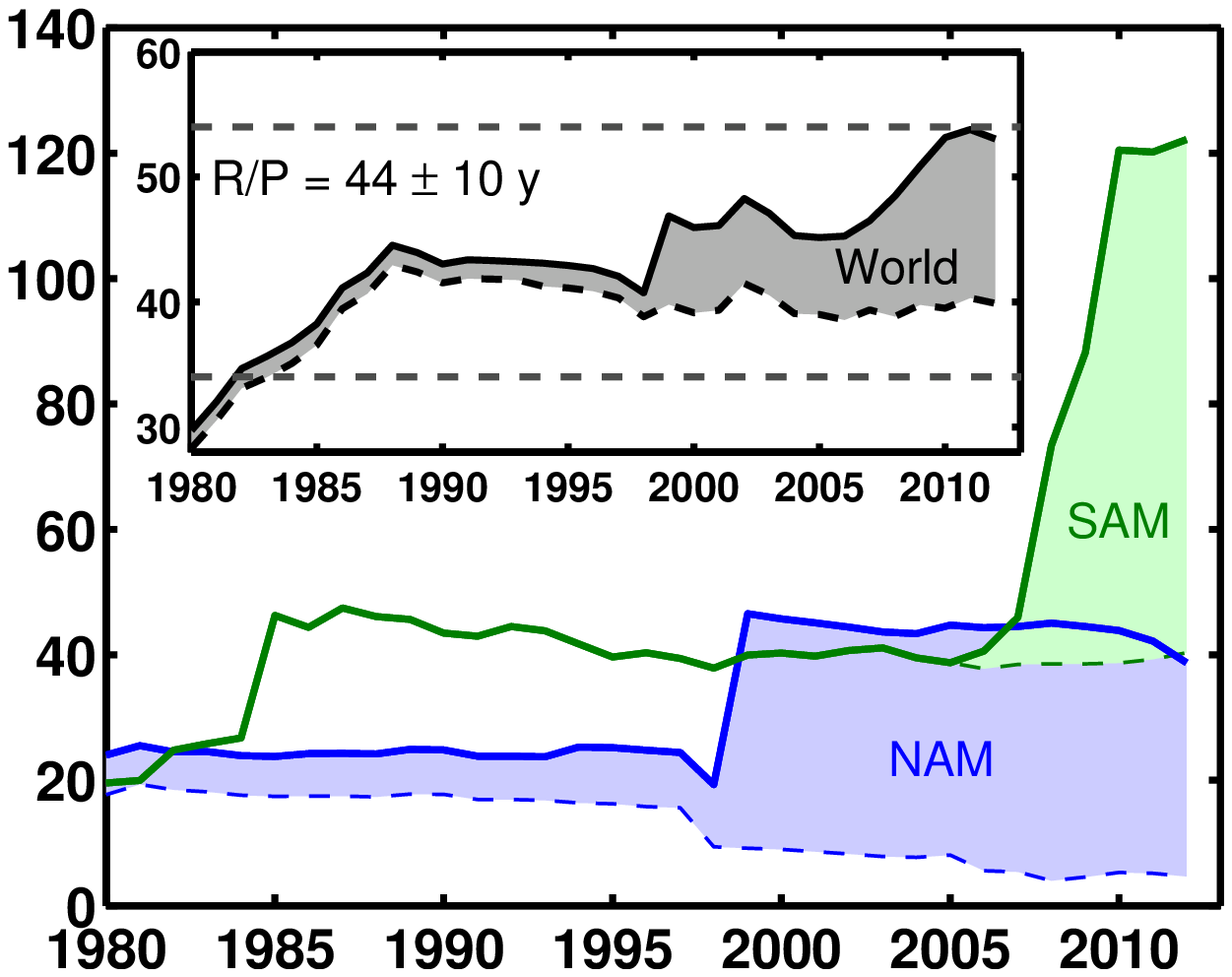}
		\end{center}
	\end{minipage}	
	\caption{Reserve to production ratios for stock resources for Oil calculated from the 2009 version of the BP Statistical Review of World Energy Workbook \citep{BP2009} (\emph{top left}) and its 2013 \citep{BP2013} version (\emph{top right}). Changes are apparent. Reserves excluding unconventional oil in the 2013 version are given in the \emph{bottom left}. Important changes in the data for NAM and SAM, and their impact on the World value, are highlighted in the panel on the \emph{bottom right}. The legend abbreviations correspond to North America (NAM), South America (SAM), Europe (EUR), Middle East (ME), Africa (AFR) and Asia-Pacific (ASP). Note how changes have been made to the data retrospectively over past years for both NAM and SAM. The dashed line on the right panel is a copy of the world total of the left panel for comparison.}
	\label{fig:Production_to_Reserve}
\end{figure}

\subsection{The determination of $\nu_0$}

These fluctuations indicate uncertainty in the determination of the $\nu_0$ parameter. We nevertheless interpret the data as indicating that $\nu_0$ \emph{is a constant of time as long as there is no change of regime for the global energy market}, but that fluctuations make its determination difficult. If no monopolistic behaviour existed in the oil market, $\nu_0$ would represent the rate at which energy reserves can be extracted from the ground. If that rate could be very fast (large $\nu_0$) in comparison to the demand, energy resources would be consumed in perfect order of cost, within an uncertainty cost range. However, even without monopolistic behaviour, since any individual oil and gas wells or coal and uranium mines can only produce at a certain finite rate (small $\nu_0$) which is lower than global demand, a number of wells and mines will remain under exploitation covering a range of costs of extraction, where expensive projects are undertaken \emph{before} resources run out in low cost projects. Adding monopolistic behaviour means increasing a local value of the R/P ratio (e.g. in ME), leading to an increase of the global value as well, indicating that some resource owners delay production while some others fill in that gap. This forces a change of regime in the world market and has to be interpreted with a change in the value of $\nu_0$. In principle, the value of $\nu_0$ need not be a constant of time; however there is no credible way by which to predict how its value could change in the future and the best approach is to keep it constant at its current value within an uncertainty range (see next section for a sensitivity analysis on changes in the value of $\nu_0$). This correspond to taking the assumption that \emph{no change of regime in the energy market} occurs in the future. The impact of the value of $\nu_0$ onto the properties of the model are described below in section \ref{sect:math}. The values for $\nu_0$ for oil are indicated in figure~\ref{fig:Production_to_Reserve} with uncertainty ranges. Note that the impact of OPEC formation is excluded from the analysis.

\section{Sensitivity analysis for the value of $\nu_0$ on future oil/gas prices and flows\label{sect:sensitivity}}

\begin{figure}[t]
	\begin{minipage}[t]{.5\columnwidth}
		\begin{center}
			\includegraphics[width=1\columnwidth]{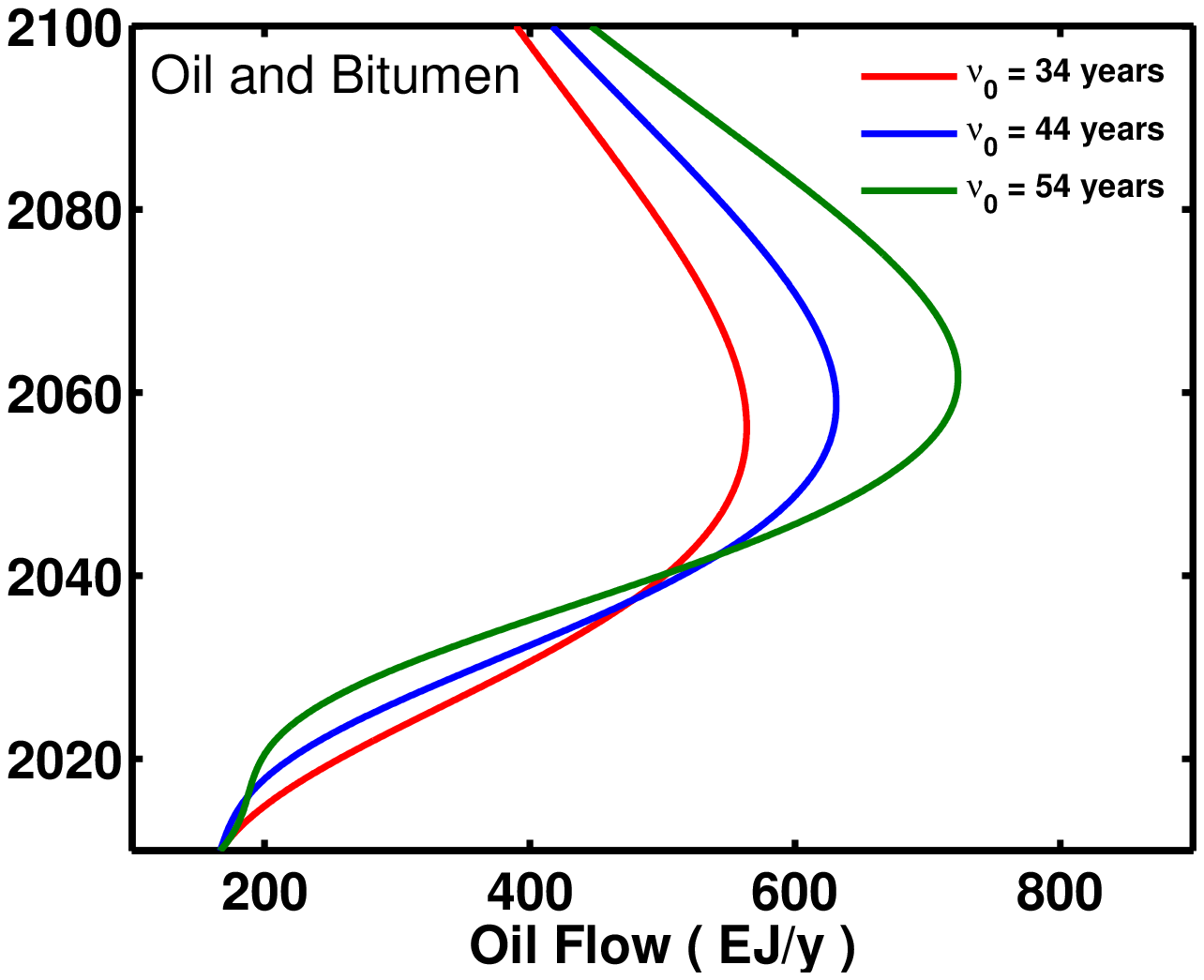}
			\includegraphics[width=1\columnwidth]{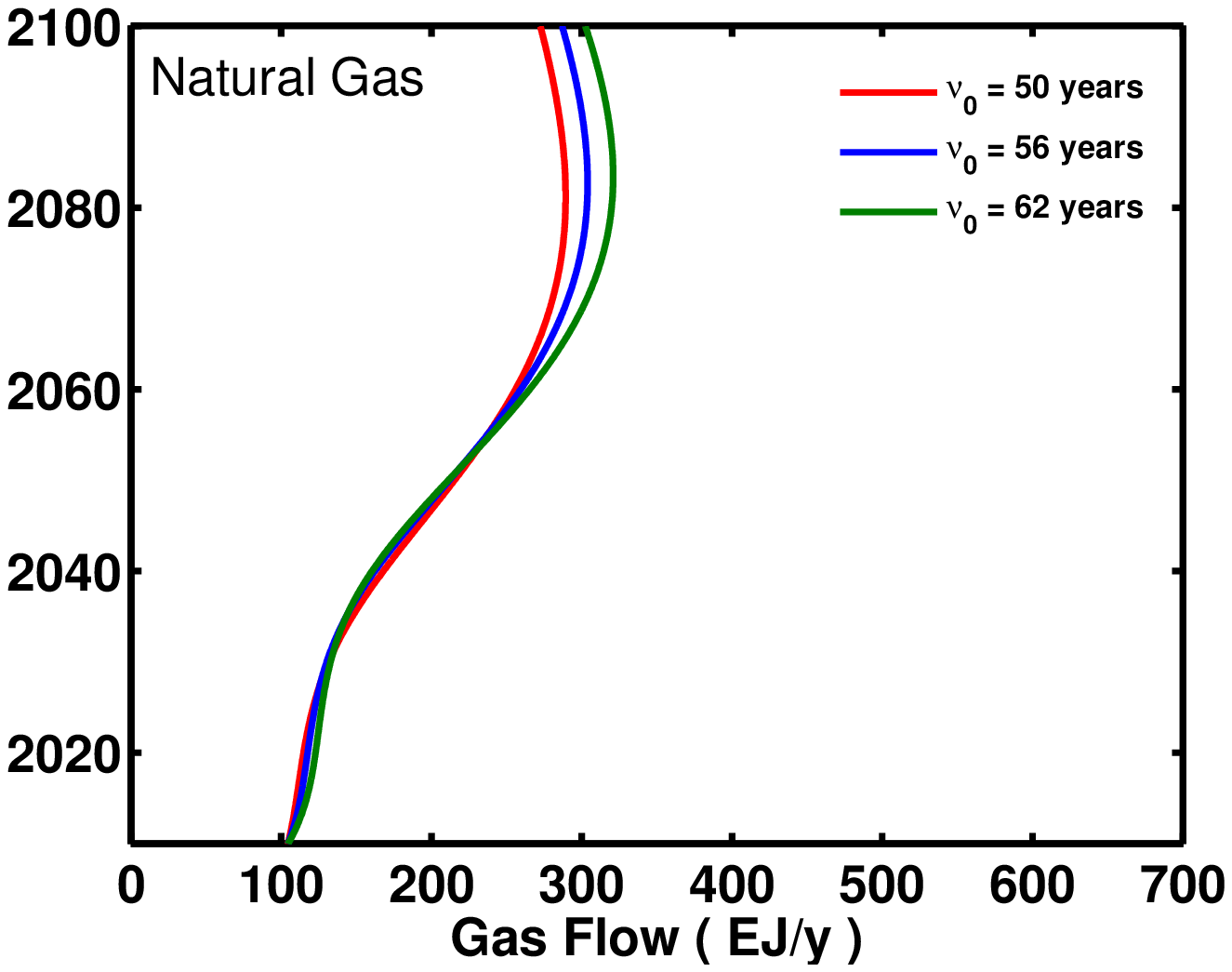}
		\end{center}
	\end{minipage}
	\hfill
	\begin{minipage}[t]{.5\columnwidth}
		\begin{center}
			\includegraphics[width=1\columnwidth]{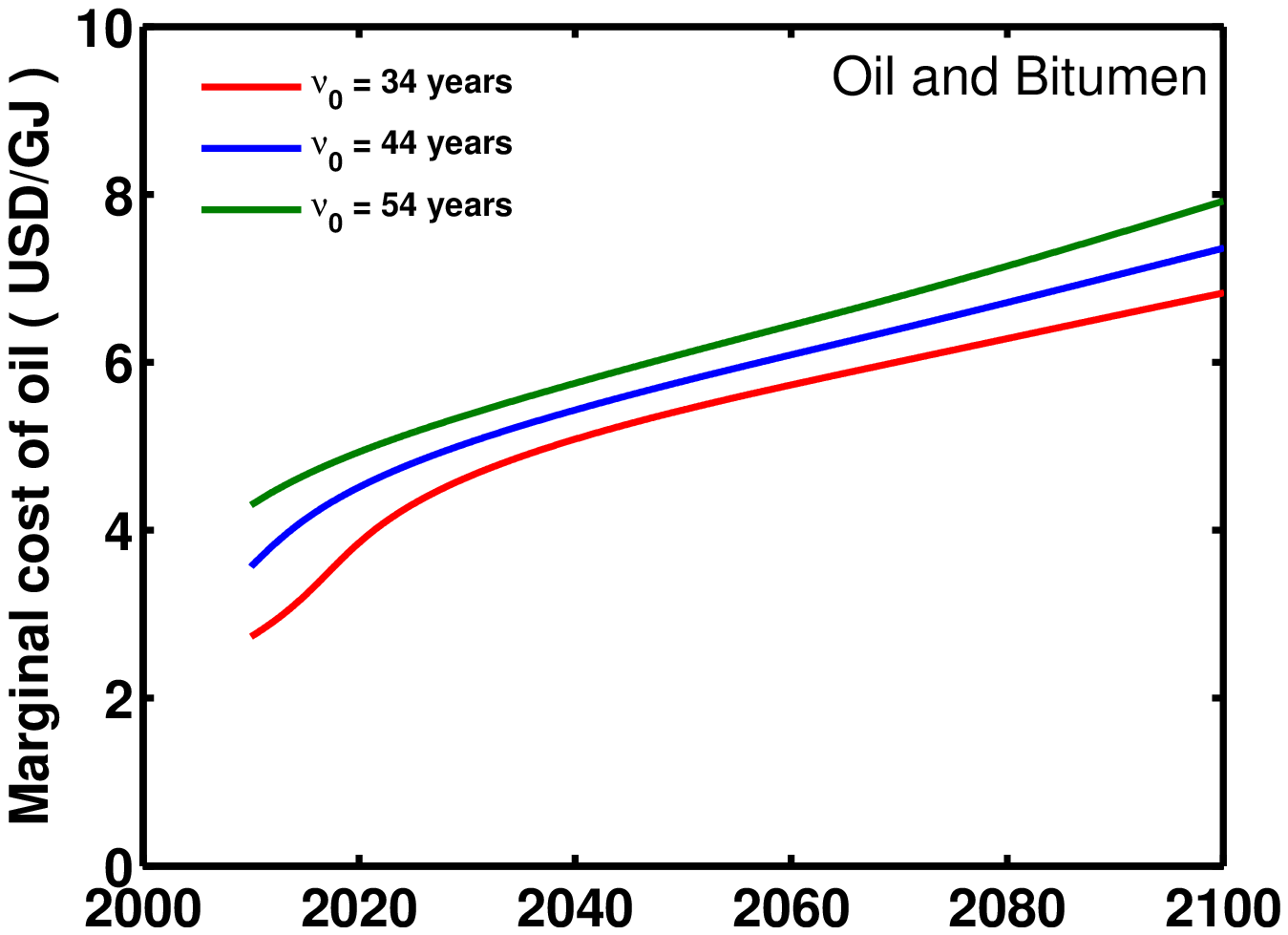}
			\includegraphics[width=1\columnwidth]{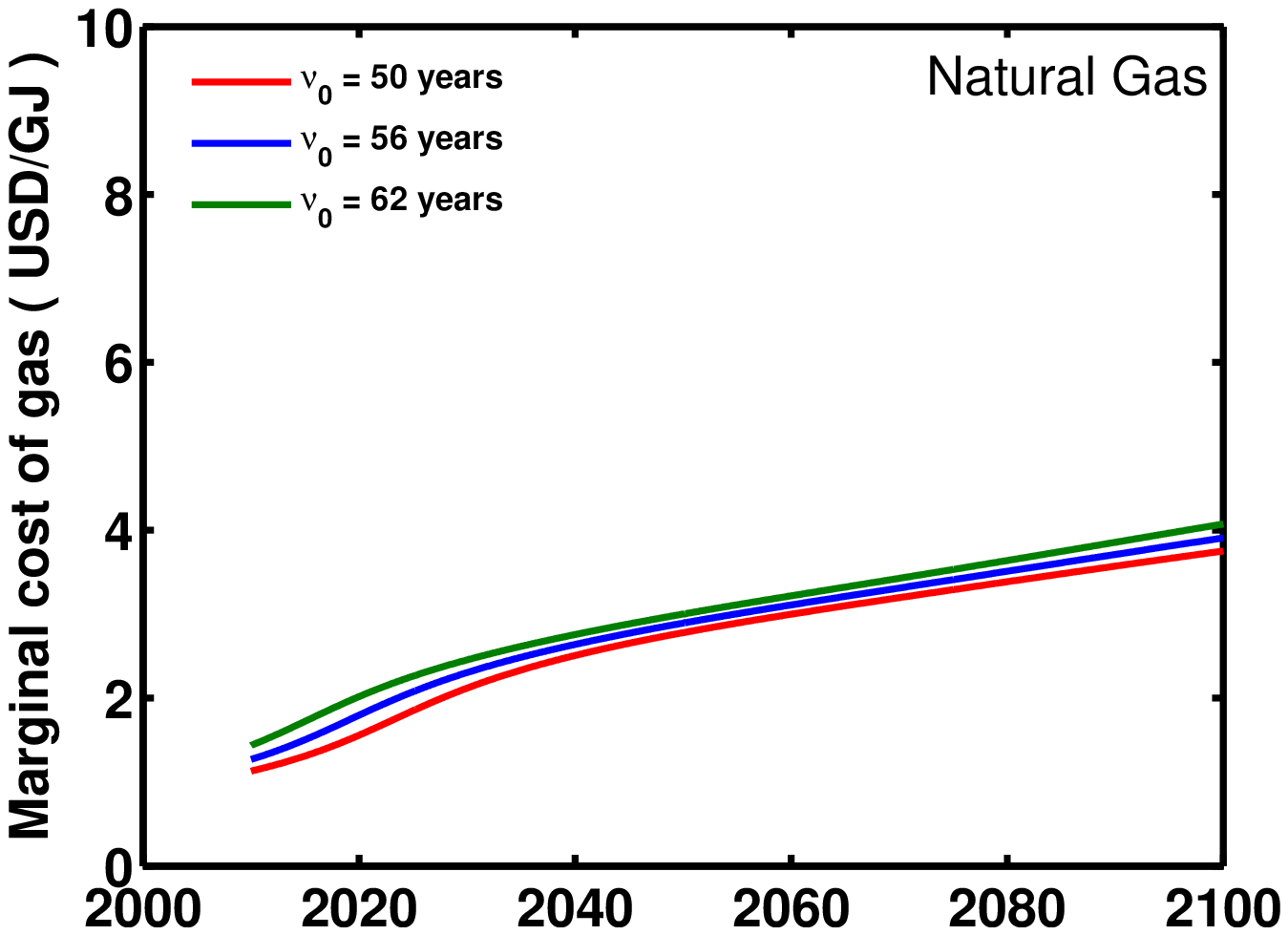}
		\end{center}
	\end{minipage}	
	\caption{(\emph{Top panels}) Sensitivity analysis for the parameter $\nu_0^{-1}$ for the oil case. In the left chart, the flow of oil is estimated using an exogenous price of oil, following the assumptions of section 3.2 of the main paper. In the right chart, the marginal cost of oil is estimated using an exogenous flow, following the assumptions of section 3.3 of the main paper. In both cases, the red, blue and green curves represent the minimum, the central and maximum values for $\nu_0^{-1}$. (\emph{Bottom panels}) Sensitivity analysis for the parameter $\nu_0^{-1}$ for natural gas.}
	\label{fig:sensitivity}
\end{figure}

According to data from \cite{BP2013} with the analysis given above, the global ratio of conventional oil reserves to production has not changed by more than 5\% during the last two decades (see Figure~\ref{fig:Production_to_Reserve}, bottom left chart). If unconventional oil is included, then the variations in the ratio increase up to 23\%, mostly due to the reclassification of Canadian tar sands and Venezuelan heavy oil from resources to reserves (see Figure~\ref{fig:Production_to_Reserve}, upper and lower right charts). In order to understand how corresponding variations in $\nu_0$ affect the results of the model, we carried out a sensitivity analysis over possible values for $\nu_0$ for oil and gas within uncertainty ranges that reflect the variations in BP data.

\subsection{Sensitivity for oil and its $\nu_0$ value}

In the model, the parameter $\nu_0^{-1}$ is considered constant and equal to 44 years for the case of oil. For the sensitivity analysis, we varied $\nu_0^{-1}$ within its uncertainty interval of 44 $\pm$ 10 years. Following the approach of sections 3.2 and 3.3 of the paper, figure~\ref{fig:sensitivity} presents the evolution in the flow of oil for the same exogenous price (left chart), and the evolution in the marginal cost of oil for the same exogenous flow (right chart). The resulting values can be directly compared to the originals of the main paper in order to estimate the level of uncertainty in comparison to other factors such as the uncertainty generated by the actual amounts of resources available. In both cases, three curves are presented, each of them corresponding to the minimum, the central and the maximum value of $\nu_0^{-1}$. 

As the left chart of figure~\ref{fig:sensitivity} shows, a rise in $\nu_0^{-1}$ delays the production of oil in the model, moving the peak from 2059 $\pm$ 3 years for minimum and maximum values of $\nu_0^{-1}$. The average flow differences are approximately of 10\%, with a maximum of 18\% in 2026. Comparing the left chart of figure~\ref{fig:sensitivity} with the upper left chart of figure 4 in the paper, it is clear that the impact in the oil flow associated to changes in $\nu_0^{-1}$ are much less important than those associated to uncertainty in the amount of oil resources. For the exogenous flow case, we can see the impact of changes in $\nu_0^{-1}$ in the marginal cost of oil in the right chart of figure~\ref{fig:sensitivity}. A rise in $\nu_0^{-1}$ increases the marginal cost of oil in 8\% average. Again, these changes are much less important than those associated to uncertainty in the amount of resources, presented in the upper left chart of figure 5 in the paper.

\subsection{Sensitivity for gas and its $\nu_0$ value}

Using the same methodology, we extended the sensitivity analysis for the natural gas industry. The parameter $\nu_0^{-1}$ presented in equation (1) of the paper is considered constant and equal to 56 years. For the sensitivity analysis, we varied $\nu_0^{-1}$ within the quoted uncertainty interval of 56~$\pm$~6 years, in accordance with the data presented in the figure~2 of the paper (right chart). Following the approach of sections~3.2 and 3.3 of the paper, figure 3 presents the flow of gas for the same exogenous price (left chart), and the evolution in the marginal cost of gas for the same exogenous flow (right chart). 

In the case of an exogenous gas price presented in figure 3 (left chart), the rise in $\nu_0^{-1}$ causes a very small delay in production, estimated at  $\pm$ 1 year for minimum and maximum values of $\nu_0^{-1}$. Regarding the flow differences, these are smaller than 6\%. For the exogenous flow case presented in figure 3 (right chart), a rise in $\nu_0^{-1}$ increases the marginal cost of gas in less than 6\%. Comparing the charts in figure~3 with figures~4 and 5 in the paper, it is clear that changes in $\nu_0^{-1}$ have much smaller impacts on the flow and price of gas than those associated to uncertainty in the amount of resources.

\section{Mathematical properties of the resource flow equation\label{sect:math}}

\subsection{Introduction}

We present in this section of the supplementary material a mathematical digression that explores the mathematical properties of equations 1 and 2 of the main text, which establish a relationship between the price $P(t)$ of an energy carrier derived from a particular type of non-renewable energy resource (e.g. coal, oil, gas, U) and its consumption, or flow, $F(t)$. As stated in the main text, the relationship is not functional, i.e. $F(t)$ cannot be written as a single valued function of $P(t)$ or the reverse. The relationship is $path$ $dependent$, and therefore depends on the history of the system and on its starting point. It thus features hysteresis. The model is slightly inspired from a physical model of energy activated processes in chemistry, as given by \cite{Mercure2005}. This material also presents the different limiting behaviours of this system for various values of its parameter $\nu_0$.

\subsection{Cost-supply curves}

\begin{figure}[t]
	\begin{minipage}[t]{.5\columnwidth}
		\begin{center}
			\includegraphics[width=1\columnwidth]{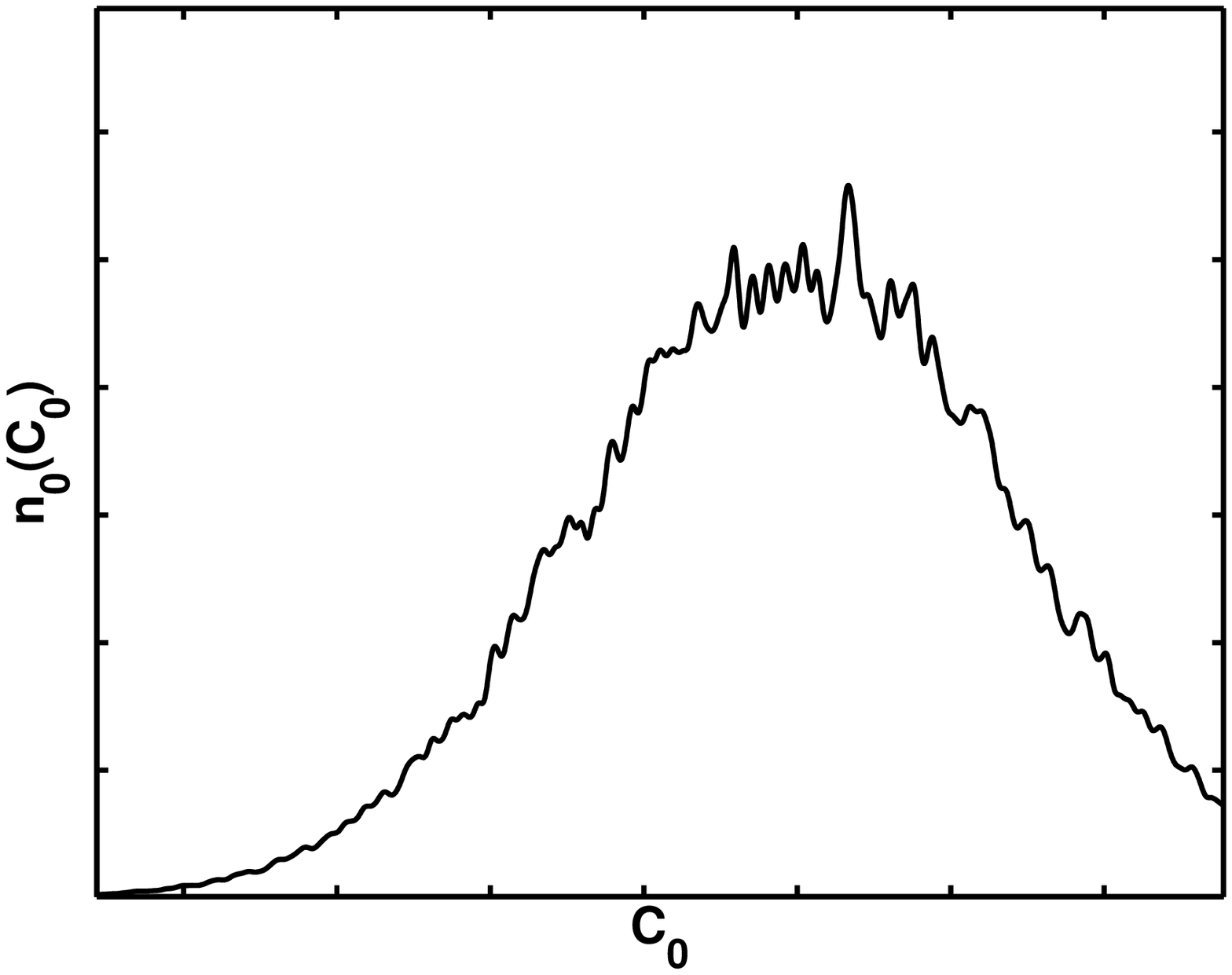}
		\end{center}
	\end{minipage}
	\hfill
	\begin{minipage}[t]{.5\columnwidth}
		\begin{center}
			\includegraphics[width=1\columnwidth]{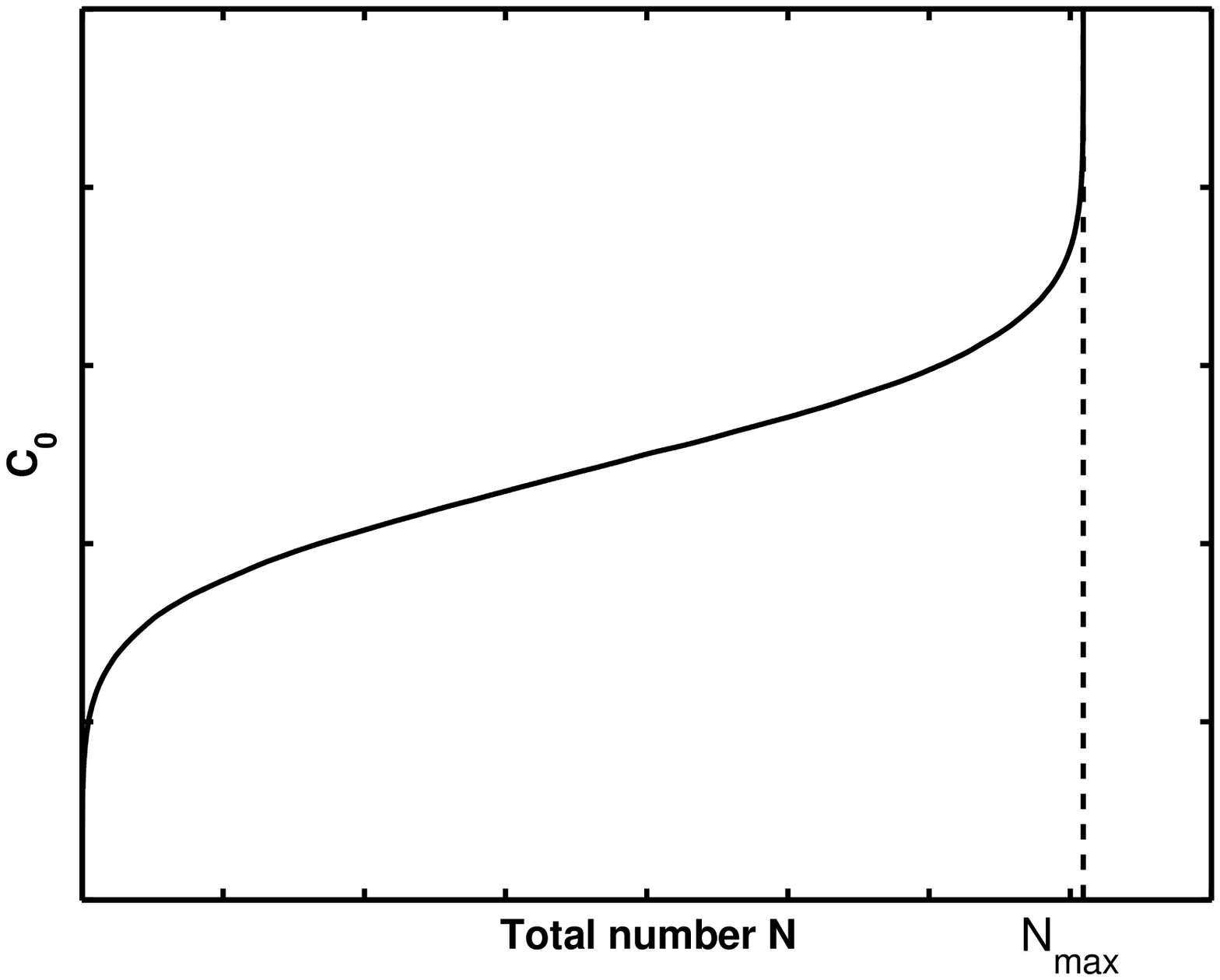}
		\end{center}
	\end{minipage}
	\caption{$Left$ Example of cost distribution for a specific natural resource $n_0(C)$ as a function of exploitation cost $C$. $Right$ Cost-quantity curve for that resource type.}
	\label{fig:OilExample}
\end{figure}

As was introduced in \cite{Mercure2012} and used in \cite{MercureSalas2012}, we define an initial (present day) distribution of non-renewable (stock) resource $n_0(C)$ function of cost $C$. This is a histogram of the number of resource units between cost values of $C$ and $C+dC$. This is a density function; the amount of resources available at costs between the values of $C_1$ and $C_2$ is
\beq
N_{1,2} = \int_{C_1}^{C_2} n_0(C)dC.
\eeq
An example of such a starting resource distribution is given in the left panel of figure \ref{fig:OilExample}. At the present day, given the cost of extraction of every resource unit, the total amount $N$ of resources available below a cost value of $C$ is
\beq
N(C) = \int_0^C n_0(C')dC'.
\eeq
Associated to this is the inverse relationship, the marginal cost of the resource $C(N)$ given that $N$ units have already been exploited; this is the cost-quantity curve. This is shown in the right panel of figure \ref{fig:OilExample}.

The amount $N$ of stock resources available below cost $C$ cannot be exploited instantaneously however, and furthermore, their owner might not be willing to extract and sell them at the current price of the associated commodity. Therefore, this amount $N$ will most likely no be sold for the price associated with the marginal cost of production $C$. If less resources are extracted at prices below $C$ than the total amount available in this cost range, resources situated higher up along the distribution must be used. Thus the cost-supply curve framework is not appropriate to use for stock resources with an international market. A similar statement could be invoked for renewable resources, however it has much less impact and the cost-supply curve framework is much more appropriate there. This is discussed in section \ref{sect:RenRes}.

\subsection{The probability of extraction}

Each resource unit has at every instant a probability of being extracted, and this probability depends on its probable individual cost of extraction $C$ and the market price of the commodity $P(t)$, $f\left(C,P(t)\right)$. Its cost of extraction is uncertain; therefore a probability distribution exists for the value of its cost of extraction $h(C)$. Additionally, the market price is stochastic and has a certain volatility, with a certain standard deviation and probability distribution $g\left(P(t)\right)$. The probability $f\left(C,P(t)\right)$ is related to $h(C)$ and $g(P)$ through the integral of their convolution, providing the `rounded step-like' probability function depicted in figure 2 of the main text. If $h$ and $g$ are normal distributions, then $f$ is an error function with width equal to the root of the sum of the squares of the widths of $h$ and $g$. These assumptions enable to avoid the use of a sharp step function, which would introduce unwanted kinks into calculations.

The probability of a resource unit of being extracted can moreover be expressed in terms of the difference between its most probable marginal cost of production and the mean value of the price, $f\left(P(t)-C\right)$.

\subsection{A differential equation for resource flows}

While the initial (present day) distribution of resources is denoted with $n_0(C)$, the distribution of future amounts of resources left as they are gradually consumed is denoted as the time dependent function $n(C,t)$. This function depends on time in two ways, by itself and through the value of the price $P(t)$, and thus strictly speaking, should be written as $n\left(C,t,P(t)\right)$. This property is the one that leads to path dependence, since the time derivative involves two terms, shown below. 

While the cost distributed amounts of resources \emph{left} is $n\left(C,t,P(t)\right)$, not all resources are exploited but only those which have a probability of being exploited, given by 
\beq
n\left(C,t,P(t)\right)f\left(P(t)-C\right).
\eeq
This corresponds to cost distributed reserves, and the size of the reserves depend on the price, expanding when the price increases. If a constant fraction $\nu_0$ of reserves are consumed with the time interval $dt$, then the flow of resources during that interval is\footnote{The fraction of cost distributed reserves consumed is actually $\nu_0 f(P(t)-C)$, a slight subtlety, which could become important is $f(P(t)-C)$ is very `rounded'.}
\beq
dn\left(C,t,P(t)\right) = -\nu_0 n\left(C,t\right)f\left(P(t)-C\right) dt,
\label{eq:main}
\eeq
which is the main equation of the model. This equation has no complete analytical solution but can be evaluated numerically, which is done for instance in FTT:Power using a discrete time step. Note that the negative sign stems from the fact that $n$ corresponds to cost distributed resources that are \emph{left} at time $t$, and that the change in resources left is negative. The flow of resources $F(t)$ is the time derivative of the total amount of resources left at all cost values $N(C,t)$, which itself is the cost integral of the resource distribution $n(C,t)$.
\beq
F\left(P(t), t\right) = - {dN \over dt} = - \int_0^\infty {dn(C,t) \over dt} dC = \int_0^\infty \nu_0 n\left(C,t\right)f\left(P(t)-C\right) dC.
\label{eq:main2}
\eeq
It is important to emphasise that the notation $F(P(t),t)$ signifies that $F$ is not a single function of  the price $P(t)$, or conversely, since an infinite number of distributions, or integrands of eq.~\ref{eq:main2}, can produce the same value for $F$.

\subsection{Mathematical properties when the price is constant}
\begin{figure}[t]
	\begin{minipage}[t]{.5\columnwidth}
		\begin{center}
			\includegraphics[width=1\columnwidth]{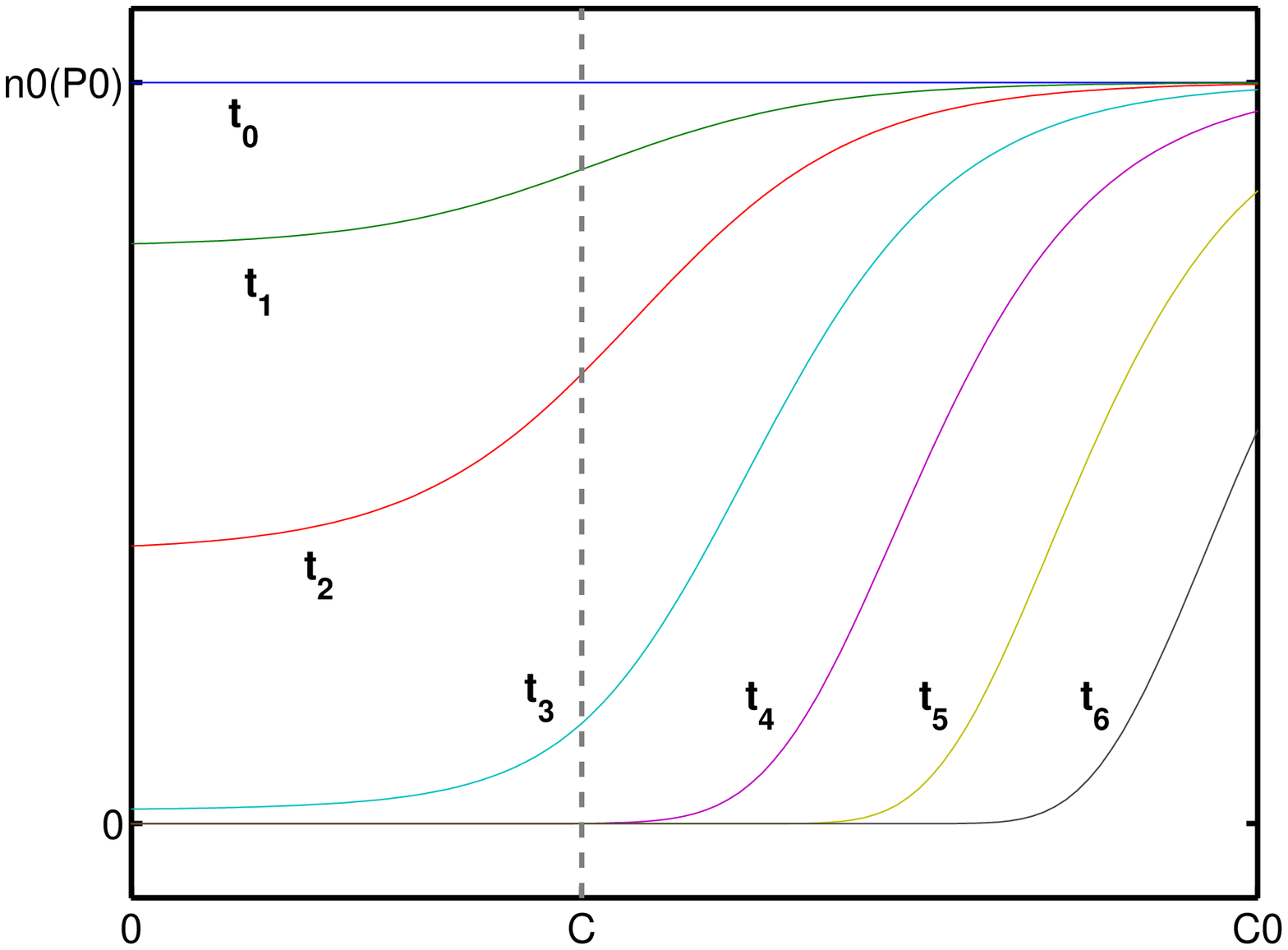}
		\end{center}
	\end{minipage}
	\hfill
	\begin{minipage}[t]{.5\columnwidth}
		\begin{center}
			\includegraphics[width=1\columnwidth]{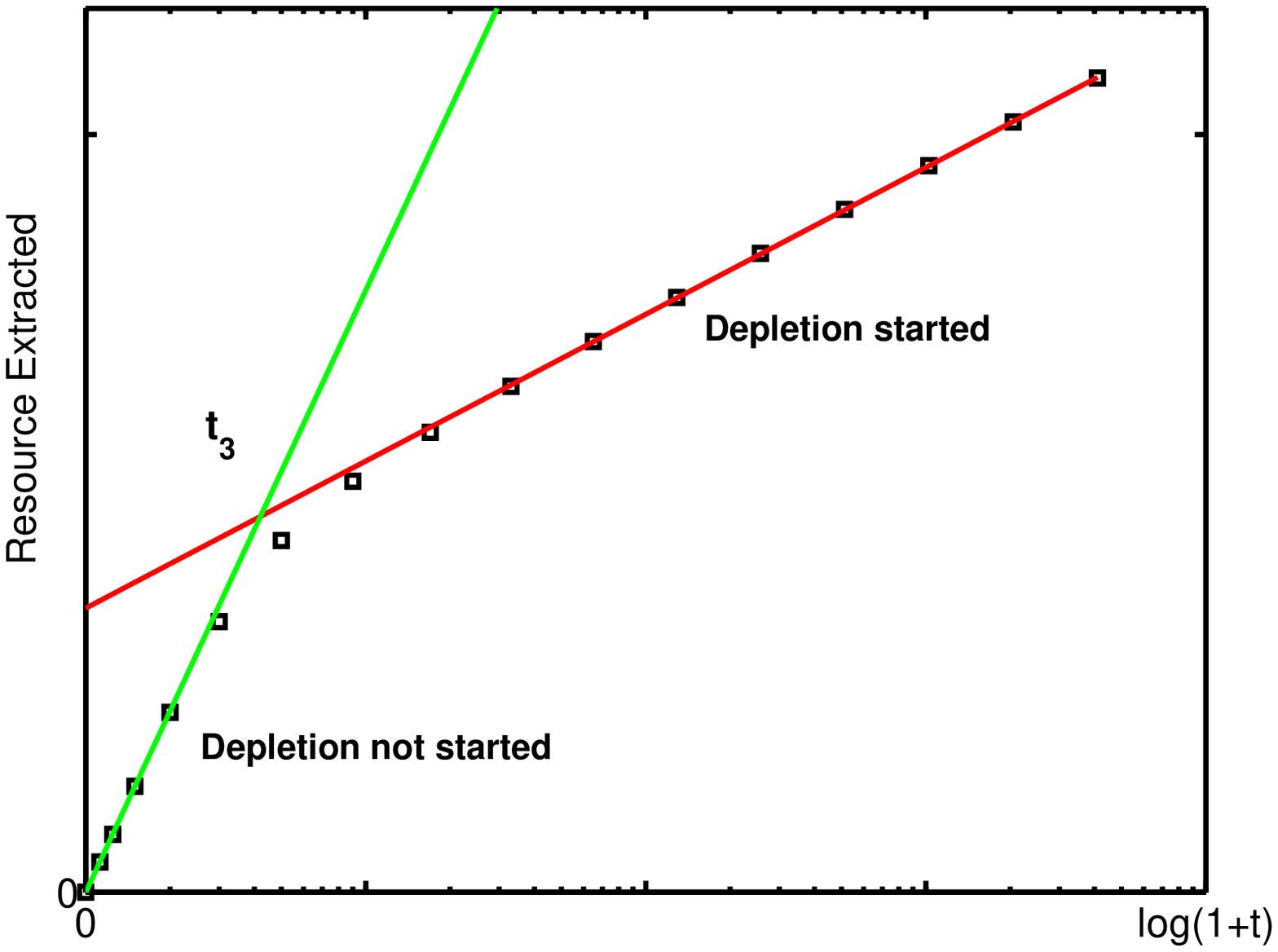}
		\end{center}
	\end{minipage}
	\caption{$Left$ Distribution of oil resources $n$ as a function of extraction cost $C$ at different times. $Right$ Total oil extracted as a function of time.}
	\label{fig:ResDep}
\end{figure}

A few important properties of equation \ref{eq:main} may be derived from the simple situation where the current price of oil $P$, is independent of time. Eq. \ref{eq:main} is solved simply:
\beq
{dn \over n} = -v_0 f(P-C) dt,
\eeq
\beq
\Rightarrow n(C,P,t) = n_0(C) e^{-t v_0 f(P-C)},
\eeq
where $n_0$ is the initial distribution of the resource previously defined, which we take, for simplicity for now, as a constant extending to very large values. We observe from this result that for a constant price, the amount of resource left at each cost value decreases exponentially in time, more and more slowly at higher and higher values of $C$.

The behaviour of $n$ as a function of $C$ is not quite as simple as this, but may be calculated numerically using a simple functional form for $f$,\footnote{A logistic function was used in this particular case; however an error function may be as appropriate.} and is shown in the left panel of fig. \ref{fig:ResDep} for different times after exploitation began. The times noted as $t_1$ to $t_6$ increase exponentially by a factor of 2 between each. Starting from the initial distribution $n_0$, a certain range of low cost units are first extracted, until they are depleted at time $t_3$. From then on, units at higher costs start to be exploited and the mid-point of $n$ starts to move towards higher values of $C_0$ up to $t_4$. $t_6$ shows $n$ after a very long time. We observe that for $t \rightarrow \infty$, all units are eventually used up and $n = 0$ for all values of $C_0$. This is due to a non-vanishing value of the probability distribution function $f(P-C)$.\footnote{The non-zero value of $f$ concerns the non-zero but very small probability that firms extract resources at a loss, due to a lack of information or miscalculations of exploitation costs.}

Furthermore, by calculating the area underneath $n_0 - n(C_0,t)$, we find the time dependence of the total number of units extracted. Fig. \ref{fig:ResDep}, right panel, shows this value against $\ln(1+t)$. We observe that before the beginning of depletion, which occurs at around time $t_3$, the extraction is fast and linear against $\ln(1+t)$. After time $t_3$, it slows down but is again perfectly linear against $\ln(1+t)$. We conclude that even though for $t \rightarrow \infty$, the system uses all existing resources, the progression of the extraction becomes exponentially slower and slower. At any practical time, the shape of $n(C_0,C,t)$ actually corresponds qualitatively to that at time $t_3$ in fig. \ref{fig:ResDep}, where low cost units have been consumed, and progresses very slowly.

Thus this demonstrates that consumption occurs even when the price does not increase. The supply, however, most likely does not meet the demand, since it monotonically decreases. 

\subsection{Consumption for a changing price}

\begin{figure}[t]
	\begin{minipage}[t]{.5\columnwidth}
		\begin{center}
			\includegraphics[width=1\columnwidth]{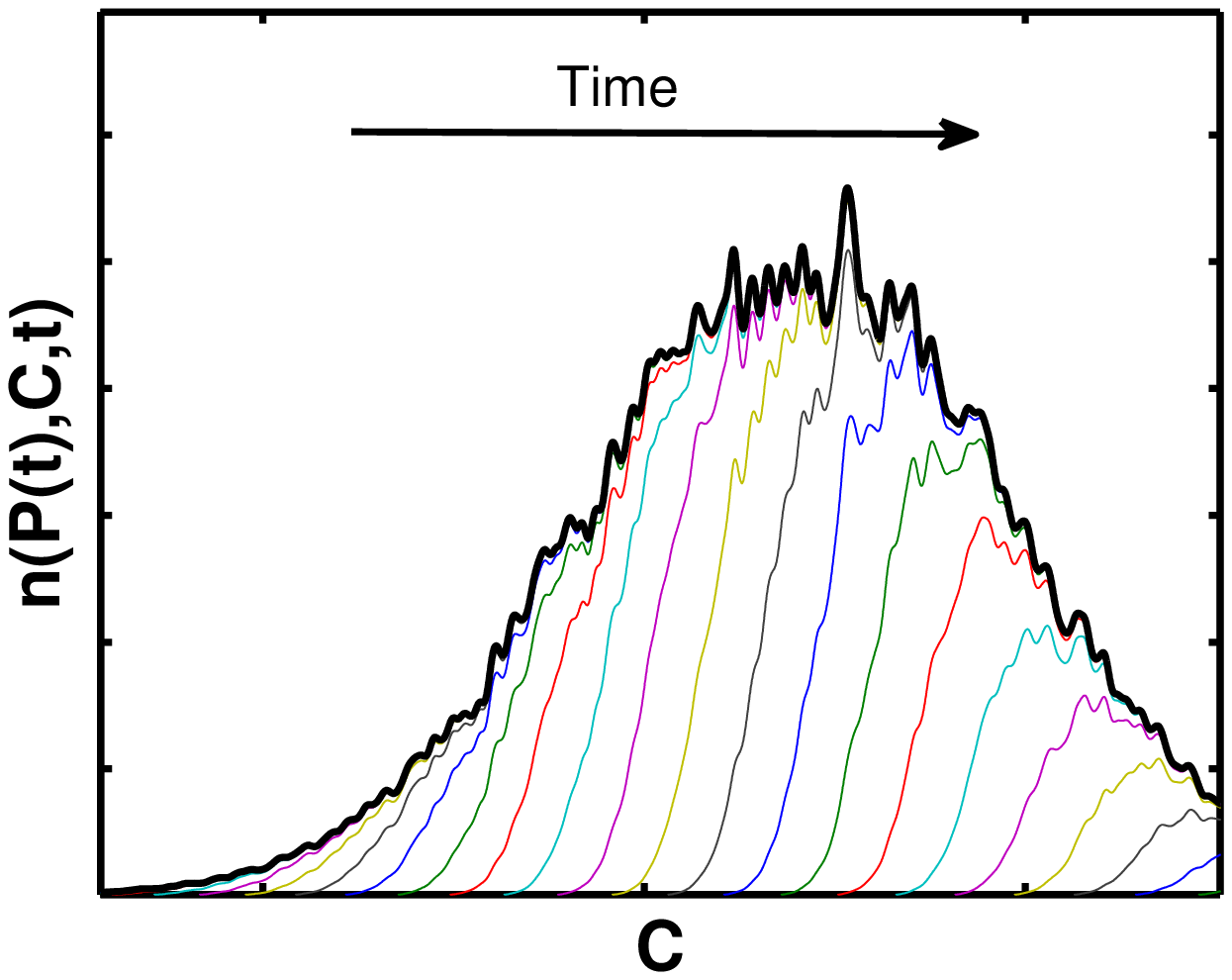}
		\end{center}
	\end{minipage}
	\hfill
	\begin{minipage}[t]{.5\columnwidth}
		\begin{center}
			\includegraphics[width=1\columnwidth]{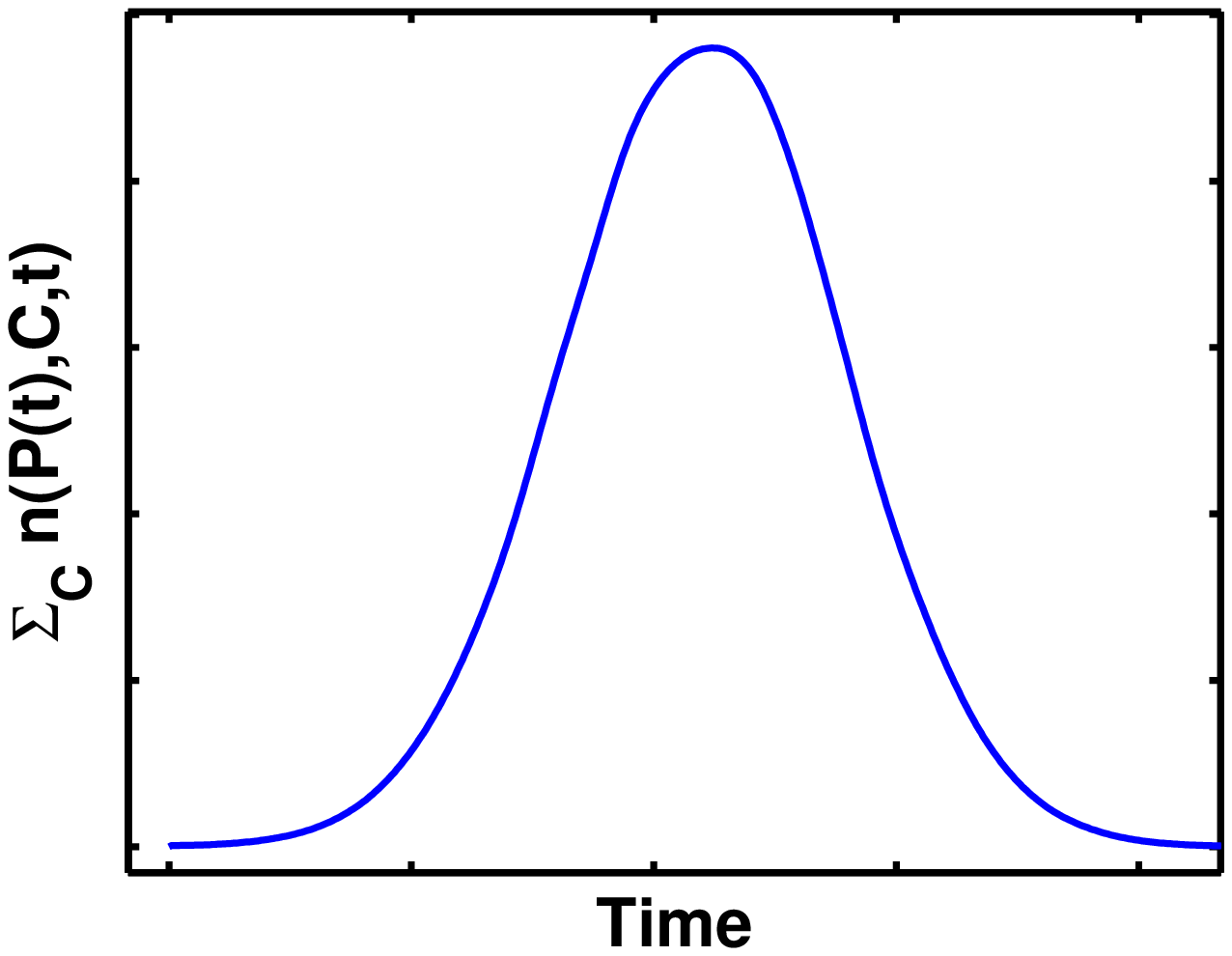}
		\end{center}
	\end{minipage}
	\caption{$Left$ Progression of the extraction of units of a resource for a linearly increasing price $P(t)$, where vertical slices of the distribution are gradually removed. $Right$ Resulting supply of units as a function of time.}
	\label{fig:ResDepTime}
\end{figure}

In the general case where the market price of units varies in time, many types of behaviour can be observed. The important mechanism at work, however, is that as low cost units are depleted, in order to keep a supply of units which does not plummet, the price of the resource must increase in time in order to make the extraction of more expensive units economical. There is thus a direct relation between supply and price, but it cannot be expressed by a simple curve. We derive here the general solution to eq. \ref{eq:main}.

The general solution to eq. \ref{eq:main} is as follows:
\beq
\int_{n_0}^{n}{dn' \over n'} = -v_0 \int_0^t f(P(t') - C) dt',
\eeq
which yields
\beq
 n(C, P(t),t) = n_0(C) e^{-v_0 \int_0^t f(P(t')-C)dt'}.
\eeq
Two time dependences are specifically denoted, that inherent to the price $P(t)$, and that associated to the exponentially decreasing exploitation that occurs at constant $C$. It is thus clear that the number of units $n(C, P(t), t)$ at any time strongly depends on the path taken by $P(t)$. The magnitude of the flow of energy extracted depends directly on the rate of change of $P(t)$. 

As an example, figure \ref{fig:ResDepTime}, left panel, depicts the progression of exploitation in time using a linearly increasing cost $P(t)$. The black curve corresponds to the initial distribution, $n_0$. Each colour curve represents the number of units as extraction progresses after a certain time. These are equally spaced in time, which results from using a linear cost progression. The right panel of the figure shows the resulting supply of units as a function of time, calculated by taking the integral of the remaining number of units $n(C,P(t),t)$ at each time value.

However, when requiring a certain supply of units, the change in $P(t)$ depends on the magnitude of $n_0(C)$ at $C$ values near that of $P(t)$. For a large numbers of units situated narrowly in cost values $C$, the value of the market cost of extraction $P(t)$ may be almost stationary. For low amounts of cheap units, the rate of change of $P(t)$ will be very large. In this theory, therefore, the value of $P(t)$ never endogenously decreases unless demand decreases. This is due to the fact that we have made abstraction of the effect of hoarding, and we thus assume that only the amount of energy intended to be consumed immediately is generated by producer firms. 

This also demonstrates the irreversibility in the process; if the price remains low, only low cost resources are used, while if the price increases, for example faster than consumption can occur, then significant amounts of low cost resources may remain in place while more expensive resources are consumed, and these two price behaviour assumptions could be chosen such that they generate exactly the same flow $F$. This corresponds to hysteresis and path dependence, akin to the change in entropy that occurs in irreversible processes in physical systems.

\subsection{The role of the parameter $\nu_0$}

\begin{figure}[t]
	\begin{minipage}[t]{.5\columnwidth}
		\begin{center}
			\includegraphics[width=1\columnwidth]{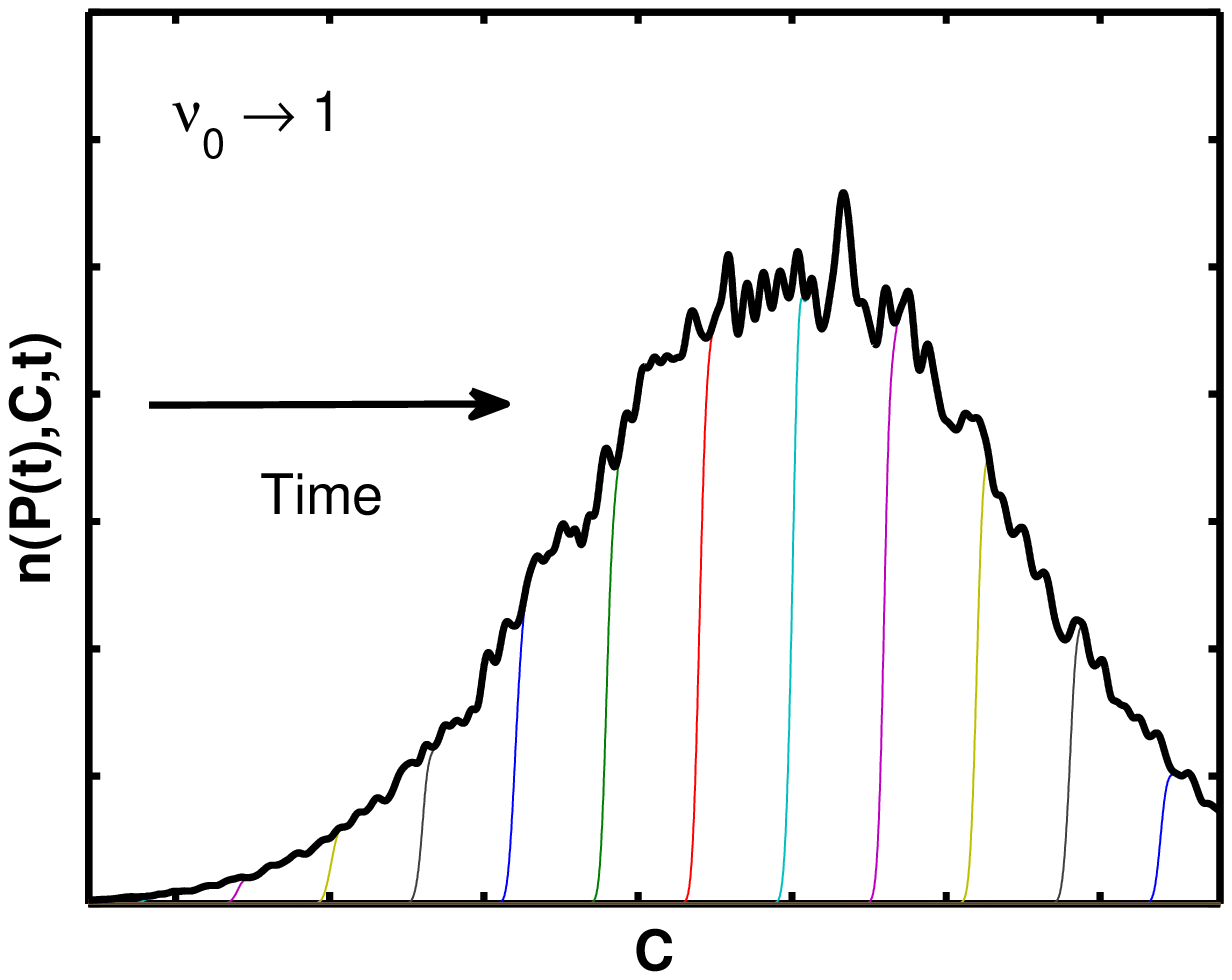}
		\end{center}
	\end{minipage}
	\hfill
	\begin{minipage}[t]{.5\columnwidth}
		\begin{center}
			\includegraphics[width=1\columnwidth]{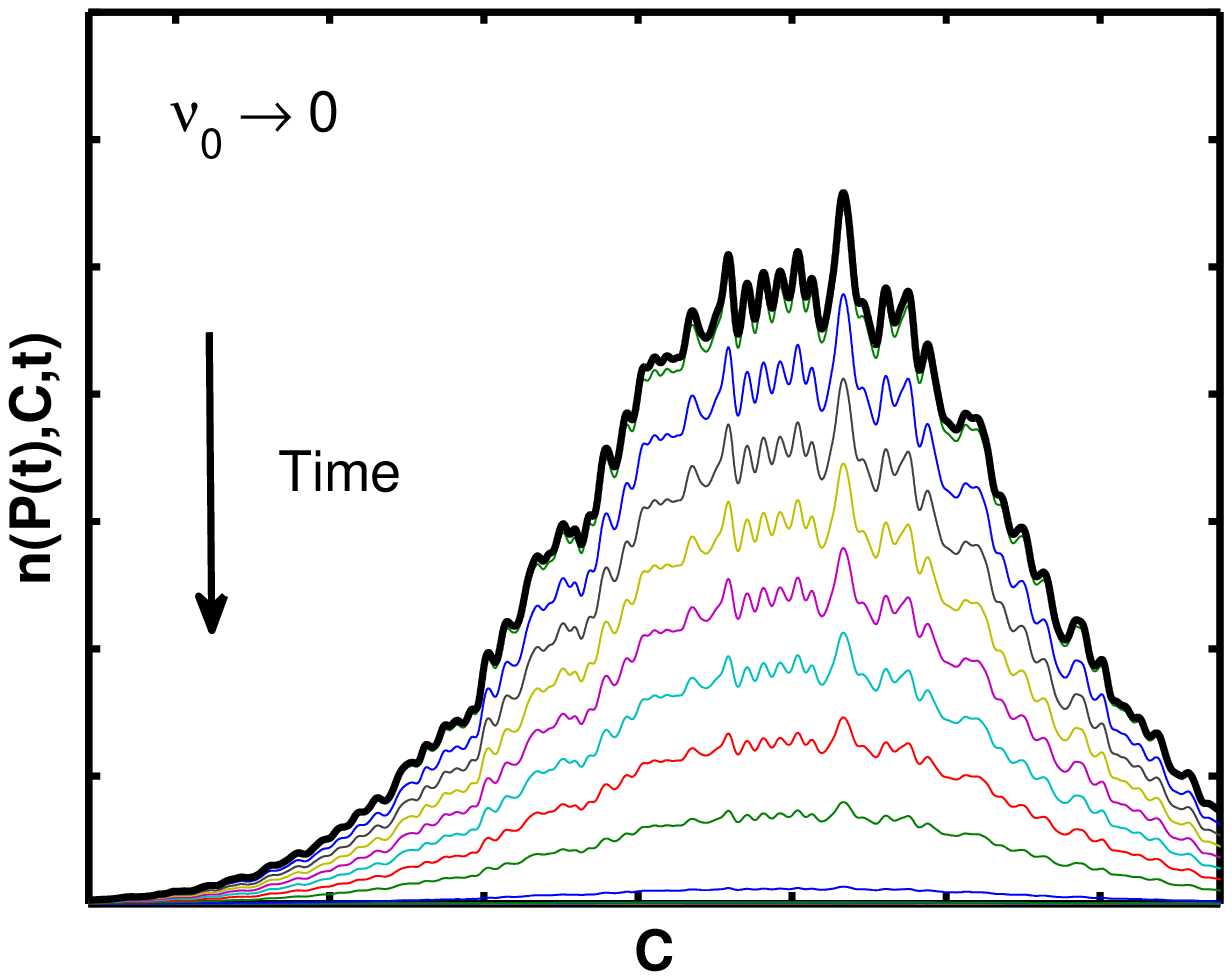}
		\end{center}
	\end{minipage}
	\caption{($Left$) Limit for $\nu_0 \rightarrow 1$. ($Right$) Limit for $\nu_0 \rightarrow \infty$.}
	\label{fig:ResDepTime}
\end{figure}

The parameter $\nu_0$ controls equation~\ref{eq:main} by determining the fraction of reserves per unit time the system is allowed to consume during a unit of time (e.g. a year), for a multitude of internal reasons that need not be known in detail. All that is required to be known is that $\nu_0$ is indeed a constant of time, a fact fairly well demonstrated for oil and gas in the main paper. But what is the effect of changing the value of $\nu_0$? This can be illustrated by the use of limiting values. 

For a value of $\nu_0 = 1$ (corresponding to a R/P ratio of 1), the complete amount of reserves can be consumed during a unit of time without requiring the price to increase. This results in the resources being exploited in perfect order of cost, with perfectly vertical slices of the distribution (as in figure \ref{fig:ResDepTime}) being consumed in order of cost. 

For a low value of $\nu_0$, the opposite behaviour occurs, where a very small fraction of reserves at each value of $C$ below $P$, producing a very low supply. This thus requires the price $P$ to increase to values high enough that the supply meets the demand. At the extreme situation where the flow does not meet the demand, the price diverges to infinity and all resources of the distribution are exploited at equal rates. The slices of figure \ref{fig:ResDepTime} thus become more or less horizontal.

Thus it can be inferred intuitively that the role of the parameter $\nu_0$ controls the size of the reserves and how high the price is required to remain in order to supply the demand. It represents to some extent the rate at which resources can physically be taken out of the ground, but also to a certain degree how willing resource rich land owners are to exploit their resources instead of keeping them for a future where they expect a higher price for them. 

\subsection{Renewable resources\label{sect:RenRes}}

This theoretical framework could also be used for renewable resources. In this case, $n(C,t)$ would correspond to a distribution of resource producing units, and the price $P(t)$ would be the price of electricity. Such a model would have quite different properties, stemming from the fundamental difference in the definition of $n(C,t)$: it concerns units of flows of resources rather than resource units. Therefore, for a constant supply of resource, i.e. a constant flow, the level of resource use does not need to change, and thus the price $P$ does not need to increase but simply to converge towards a constant value. This therefore shows that this model used for renewables would be very close to equivalent to a cost-supply curve framework, where a supply corresponds closely to a single cost value. There would be little gain in attempting to define such a model for renewables, and thus this is not done in the FTT model.

\end{document}